\documentclass{aa}  
\usepackage{graphicx}
\usepackage{multirow}
\usepackage{xcolor}
\usepackage[caption=false]{subfig}
\usepackage{bm}
\usepackage{txfonts}
\usepackage[colorlinks=true, allcolors=blue]{hyperref}

\usepackage{lscape}

\usepackage{soul}

\begin{document} 

    \title{Evolution of the relation between the mass accretion rate and the stellar and disk mass from brown dwarfs to stars
    \thanks{Based on observations collected at the European Southern Observatory under ESO programme     105.20NP.001}
    }
    \titlerunning{}
    \authorrunning{Almendros-Abad et al.}
    

   \author{ V. Almendros-Abad \inst{1}, C. F. Manara\inst{2},   L. Testi\inst{3}, A. Natta\inst{4}, R. A. B. Claes\inst{2}, K. Mu\v{z}i\'c\inst{5}, E. Sanchis\inst{6}, J. M. Alcalá\inst{7}, A. Bayo\inst{2}, \and A. Scholz\inst{8}
          }

        \institute{CENTRA, Faculdade de Ci\^{e}ncias, Universidade de Lisboa, Ed. C8, Campo Grande, P-1749-016 Lisboa, Portugal\\
        \email{victor.almendrosabad@inaf.it}
        \and European Southern Observatory, Karl-Schwarzschild-Strasse 2, 85748 Garching bei München, Germany
        \and Alma Mater Studiorum Università di Bologna, Dipartimento di Fisica e Astronomia (DIFA), Via Gobetti 93/2, 40129 Bologna, Italy
        \and School of Cosmic Physics, Dublin Institute for Advanced Studies, 31 Fitzwilliam Place, Dublin 2, Ireland
        \and Instituto de Astrofísica e Ciências do Espaço, Faculdade de Ciências, Universidade de Lisboa, Ed. C8, Campo Grande, 1749-016 Lisbon, Portugal
        \and Freie Universität Berlin, Institute of Geological Sciences, Malteserstr. 74-100, 12249 Berlin, Germany
        \and
        INAF-Osservatorio Astronomico di Capodimonte, Via Moiariello 16, 80131 Napoli, Italy
        \and
        SUPA, School of Physics \& Astronomy, University of St Andrews, North Haugh, St Andrews, KY16 9SS, United Kingdom
        }

   \date{Received; accepted}

  \abstract
   {The time evolution of the dependence of the mass accretion rate with the stellar mass and the disk mass represents a fundamental way to understand the evolution of protoplanetary disks and the formation of planets. In this work, we present observations with X-Shooter of 26 Class II very low-mass stars ($<$0.2 $M_\odot$) and brown dwarfs in the Ophiuchus, Chamaeleon-I, and Upper Scorpius star-forming regions. These new observations extend down to the spectral type (SpT) of M9 ($\sim$0.02 $M_\odot$), which is the measurement of the mass accretion rate in Ophiuchus and Chamaeleon-I and add 11 very-low-mass stars to the sample of objects previously studied with broadband spectroscopy in Upper Scorpius. We obtained the spectral type and extinction, as well as the physical parameters of the sources. We used the intensity of various emission lines in the spectra of these sources to derive the accretion luminosity and mass accretion rates for the entire sample. Combining these new observations with data from the literature, we compare relations between accretion and stellar and disk properties of four different star-forming regions with different ages: Ophiuchus ($\sim$1 Myr), Lupus ($\sim$2 Myr), Chamaeleon-I ($\sim$3 Myr), and Upper Scorpius (5-12 Myr). We find the slopes of the accretion relationships ($L_*-L\mathrm{_{acc}}$, $M_*-\dot{M}\mathrm{_{acc}}$) to steepen in the 1-3 Myr age range (i.e., between Ophiuchus, Lupus, and Chamaeleon-I) and that both relationships may be better described with a single power law. We find that previous claims for a double power-law behavior of the $M_*-\dot{M}\mathrm{_{acc}}$ relationship may have been triggered by the use of a different SpT-$T\mathrm{_{eff}}$ scale. We also find the relationship between the protoplanetary disk mass and the mass accretion rate of the stellar population to steepen with time down to the age of Upper Scorpius. Overall, we observe hints of a faster evolution into low accretion rates of low-mass stars and brown dwarfs. At the same time, we also find that brown dwarfs present higher $M\mathrm{_{disk}}/\dot{M}\mathrm{_{acc}}$ ratios (i.e., longer accretion depletion timescales) than stars in Ophiuchus, Lupus, and Cha-I. This apparently contradictory result may imply that the evolution of protoplanetary disks around brown dwarfs may be different than what is seen in the stellar regime.}


   \maketitle
%

\section{Introduction}
\label{intro}

The evolution of protoplanetary disks ultimately sets the conditions for planet formation. During this process, protoplanetary disks drastically change their dust and gas distributions as well as their morphology. In order to understand planet formation, obtaining a better grasp of how this evolution takes place is fundamental. The evolution of protoplanetary disks is driven by the following physical processes: accretion of material through the disk onto the central star \citep{hartmann16}, ejection of material through winds \citep{pascucci22}, and external factors such as binarity, encounters \citep{cuello23}, or external photoevaporation \citep{winter22,mauco23}, which can also signficantly contribute to the loss of material in protoplanetary disks.

A very practical way to study the evolution of disks is to measure their bulk properties for a large number of young stellar objects in star-forming regions of different ages, and spanning a wide range of stellar masses. Strong efforts have been made in the last decade to measure the mass accretion rate \citep[$\dot{M}\mathrm{_{acc}}$;][]{alcala14,alcala17,manara15,manara17cha,manara20} and the disk mass \citep[$M\mathrm{_{disk}}$;][]{barenfeld16,ansdell16,pascucci16,testi16,sanchis20,miotello23} of young stellar objects, using instruments such as X-Shooter and ALMA, respectively. Some of the most important results based on the relationship between the measurements of these bulk properties are: a steep correlation between  the stellar mass ($M_{\mathrm{*}}$) and $\dot{M}\mathrm{_{acc}}$ \citep{hillenbrand92,mohanty05,natta06} and between $M\mathrm{_{disk}}$ and $M\mathrm{_*}$ \citep{mohanty13,andrews13,barenfeld16,pascucci16,rilinger23}. The $M\mathrm{_{disk}}$ and $\dot{M}\mathrm{_{acc}}$ are also correlated and have a shallower slope close to linear \citep{manara16,mulders17}. There is a decline of the disk mass and steepening of its relationship with stellar mass as a function of time \citep{ansdell17,pascucci16}. The relation between these properties and the stellar ones can provide very strong tests for disk evolution and planet formation models \citep[see][and references therein]{manara23}. 

Focusing on the low-mass star and brown dwarf (BD) side of these relationships, there have been several intriguing results: it was found that the $M_{\mathrm{*}}$--$\dot{M}\mathrm{_{acc}}$ relationship of the Lupus \citep{alcala17} and Chamaeleon-I \citep[Cha-I hereafter,][]{manara17cha} star-forming regions ($\sim$2-3 Myr) may experience a steeper decrease of the accretion rates in the very low-mass star regime ($<$0.2-0.3 M$_\odot$) than in the solar-mass regime. On the other hand, in the younger Ophiuchus and NGC 1333 star-forming regions ($\sim$1 Myr), the entire population down to the star-BD boundary follows the same scaling relation \citep{manara15,testi22,fiorellino21}, which could suggest that this is an evolutionary effect, where low-mass stars have a faster decline of their accretion rates. Measurements of the accretion rate in free-floating BDs appear to suggest that BDs follow the accretion behavior of low-mass stars \citep{mohanty05,muzerolle05,herczeg08,alcala17,manara15,manara17cha}, whereas there is tentative evidence that planetary-mass companions do follow the predictions of \citet{stamatellos15} and present a mass-independent accretion rate \citep[e.g.,][]{betti23}. However, \citet{betti23} recently found that BDs may be better characterized by a steeper slope than stars in the $M_{\mathrm{*}}$--$\dot{M}\mathrm{_{acc}}$ relationship and that this relationship steepens with increasing age. When compared with the disk dust mass, \citet{sanchis20} found that BDs in the Lupus star-forming region have larger $M\mathrm{_{disk}}/\dot{M}\mathrm{_{acc}}$ ratios than stars (i.e., longer accretion depletion timescale), indicating that BDs present lower accretion rates for the same disk masses. There is also some evidence that protoplanetary disks around BDs may be longer-lived than those around their higher mass counterparts \citep{scholz07,luhman12usco}. All these issues may have a profound implication in our understanding of how BDs form and  their ability to form their own planetary systems. However, samples of BDs with available accretion and protoplanetary disk masses measurements are currently very limited \citep[$<$50\% completeness at spectral type, SpT) later than M6 in all regions where the accretion rate has been studied extensively,][]{alcala17,manara17cha,manara15,natta06,testi22}.

At the same time, disks around BDs present several similarities to those around higher-mass stars: they undergo a T Tauri-like phase during their early evolution where they accrete material from the disk and produce outflows \citep{jayawardhana03,natta04,whelan05}, they undergo dust processing and grain growth \citep{sterzik04,apai05,pinilla17}, their disks are typically flat \citep{scholz07}, and some show signs of inner disk clearing \citep{muzerolle06,rilinger19}. This similarity between disks around BDs and stars hints that the process for the formation of BDs can be consistent with that of stars. It also tells us that BDs may also be able to produce their own planetary systems, given that protoplanetary disks around BDs and very-low-mass stars ($<$0.2 $M_\odot$) present gaps in their mass distribution like their higher mass counterparts \citep{kurtovic21,pinilla21}. However, a planetary origin of these gaps pose a strong challenge for planet formation models, since they would require a planet with at least a Saturn mass \citep{pinilla17,sinclair20}, which is very challenging for models of planet formation to reproduce \citep{miguel20,mercer20,morales19}.

This work presents the study of the relationships between the stellar, accretion and disk observables in four different star-forming regions: Ophiuchus, Lupus, Cha-I, and Upper Scorpius. The goal of this work is to study whether the evolution of the protoplanetary disk mass accretion rates of young stellar objects is the same across all masses. In the context of BDs, understanding whether they follow low-mass stars in the evolution of the scaling relations between the stellar and disk observables can provide fundamental information about the physical processes in protoplanetary disks around BDs. This work benefits from new observations with VLT/X-Shooter of 26 low-mass stars and BDs in Ophiuchus, Cha-I, and Upper Scorpius (12 of which have SpT$\geq$M6). In Sect.~\ref{xshoo_observations}, we present the X-Shooter observations, their data reduction, and literature measurements of the accretion rate and the protoplanetary disk mass. In Sect.~\ref{xshoo_analysis}, we analyze the X-Shooter observations, including the derivation of the SpT, extinction, and the accretion luminosity. In Sect.~\ref{xshoo_results}, we present the accretion and disk scaling relations ($L_*-L\mathrm{_{acc}}$, $M_*-\dot{M}\mathrm{_{acc}}$, $M_*-M\mathrm{_{disk}}$, $M\mathrm{_{disk}}-\dot{M}\mathrm{_{acc}}$) for the four star-forming regions studied in this work and their analysis. In Sect.~\ref{xshoo_discussion} we discuss these results and place them in the context of the understanding of disk evolution. Lastly, in Sect.~\ref{xshoo_summary} we summarize our findings and present the conclusions of this work.


\section{Observations and data reduction}
\label{xshoo_observations}

The observed sample was selected to extend the accretion rate measurements in the Ophiuchus youngest cloud L1688, the Upper Scorpius, and the Cha-I star-forming regions to lower masses. The sample was originally selected from the members of these regions  with available or scheduled ALMA observations \citep{carpenter14,barenfeld16,testi16,testi22,cieza19}. However, the ALMA program in Cha-I was not completed, while some of the Ophiuchus targets observed in this work had to be changed for sources without ALMA observations because observations were unfeasible due to a lack of available guide stars. All of the observed sources presented infrared excess, found using \textit{Spitzer} and \textit{Wise} \citep{luhman12usco,esplin20,luhman08}. The final sample includes 10 low-mass stars and BDs from Ophiuchus, 11 low-mass stars from Upper Scorpius, and 5 BDs from Cha-I. All the targets were observed with the X-Shooter \citep{vernet11} spectrograph at the VLT, under program ID 105.20NP.001 (PI E. Sanchis). X-Shooter provides simultaneously a wide spectral coverage (310-2500 nm), with a moderate spectral resolution. The spectrum is divided in three arms that are usually referred to as the UVB ($\lambda\lambda\sim$ 310-550 nm), VIS ($\lambda\lambda\sim$ 550-1050 nm), and near-infrared (NIR: $\lambda\lambda\sim$ 1050-2500 nm) arms.

In Table~\ref{tab:tab_obs} we provide the log of the X-Shooter observations. The targets from Upper Scorpius (brightest in the sample) were observed with the 1.0$''$, 0.4$''$, and 0.4$''$ slits, leading to a spectral resolution R $\sim$ 5400, 18400, 11600 in the UVB, VIS, and NIR arms, respectively. The rest of the targets were observed using the 1.0$''$, 0.9$''$, and 0.9$''$ slits leading to a spectral resolution R $\sim$ 5400, 8900, and 5600. The observations with these narrow slits were carried out in nodding ABBA to allow for proper sky subtraction. Prior to the narrow slit observations, each target was also observed using wide slits in the three arms (5$''$ wide). These wide slit spectra were used to perform the absolute flux calibration as in \citet{alcala17,manara17cha}.

The data reduction was carried out using the X-Shooter pipeline v.3.5.3 \citep{modigliani10}, which included bias and flat-field corrections, order extraction and combination, rectification, wavelength calibration, relative flux calibration using standard stars observed in the same night, and extraction of the spectrum. Telluric correction was performed using the \textit{molecfit} tool \citep{kausch15,smette15} v.4.2.1.9 for the VIS and NIR arms. The correction was performed by fitting the atmospheric model directly on the spectra except for three cases (2MASS J11112249-7745427, CRBR 2317.5-1729, and BKLT J162736-245134), where we used the telluric standard star observed in the same night. We corrected the spectra for slit losses using the wide slit spectra observed right before the narrow slit observations. In the UVB arm we used a constant correction factor (median between the flux of the wide and narrow slit spectra) for all the targets. In the VIS arm we used a constant factor for the targets that were observed with the 0.9$''$ slit, while for the rest of the targets we applied a wavelength-dependent correction factor. In the NIR arm, we adopted a wavelength-dependent correction for all the targets. We visually evaluated that there is a smooth transition between the different arms of the flux calibrated spectra. We observed that some of the targets with low signal in the VIS arm present a jump in flux in the VIS-NIR transition. For these objects (CRBR 2317.5-1729, CRBR 2322.3-1143, GY92 90, ISO-Oph042, and SONYC RhoOph-6), we applied a constant correction factor to the VIS arm spectra to match the blue end of the NIR arm spectra. None of these objects presented any signal in their UVB arm spectrum.

The wide slit observations of EPIC 203756600 were obtained with the 1.6$''$, 1.5$''$, 1.2$''$ slits in the three arms, respectively, in order to avoid contamination by a brighter T Tauri star closer than 5$''$ (which is also a member of Upper Scorpius). By comparing with archival photometry we observe that only the NIR arm suffered from some slit losses (maximum seeing during the observations of 1.6$''$), therefore we rescaled the NIR arm to match the red end of the 1.5$''$ observations in the VIS arm with a constant scaling factor.

We compared the flux-calibrated spectra with archival photometry for all the targets and found them to be in agreement within the typical flux uncertainty of 5-10\%. We find GY92 264 to be $\sim$35\% fainter than the photometric flux, as well as the telluric standard star observed the same night (there were thin cirrus clouds present during the observations). We performed a constant factor flux correction of the target spectrum based on the comparison of the telluric standard star with its photometric fluxes.


\section{Analysis} 
\label{xshoo_analysis}

\subsection{Derivation of spectral type and extinction}
\label{analysis_templates}

\begin{figure}[hbt!]
    \centering
    \includegraphics[width=\textwidth/21*10]{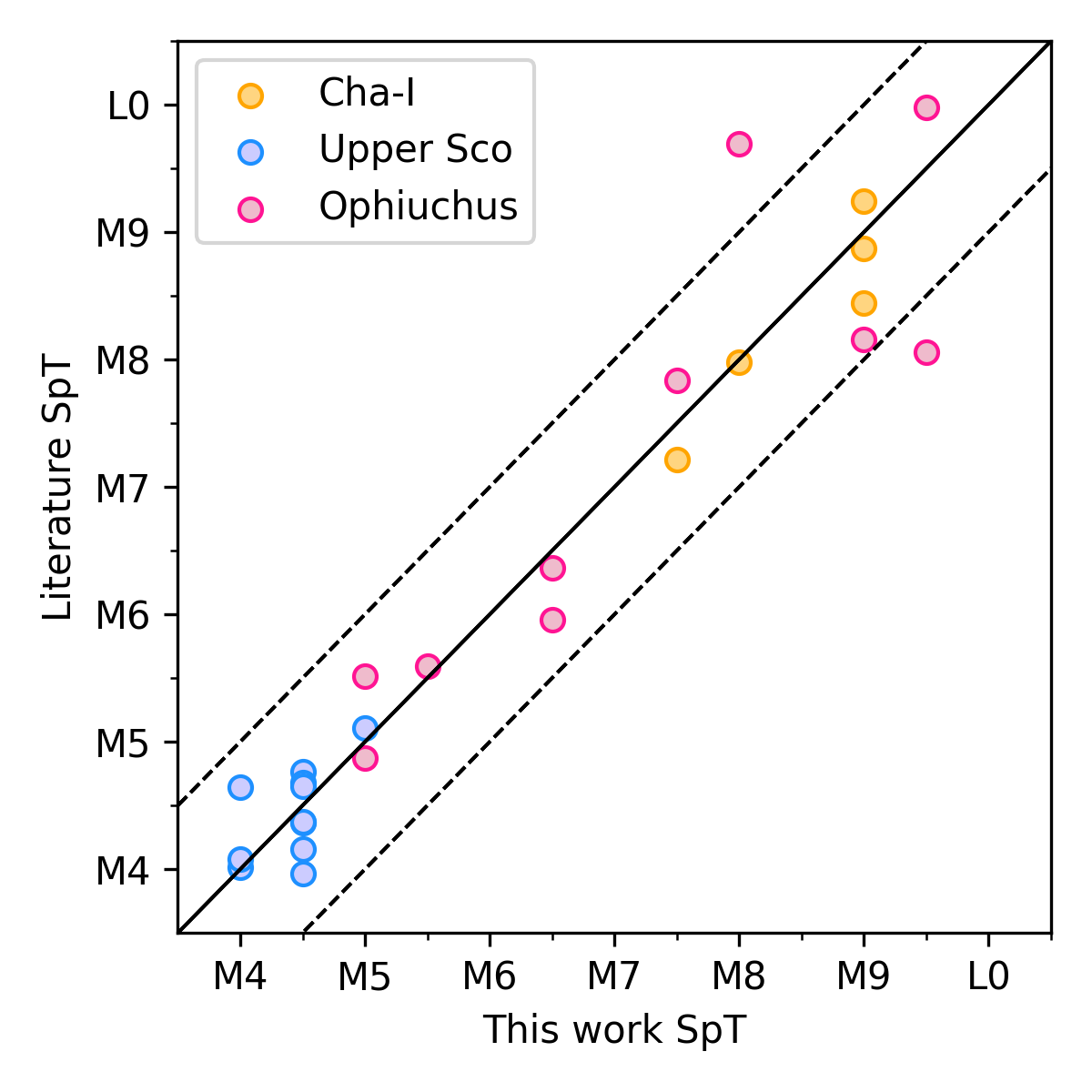}
    \caption{Comparison of the SpT measured in this work (see Sect.~\ref{analysis_templates}) and the SpT from the literature. Sources belonging to Cha-I are shown with orange circles, to Upper Scorpius with blue circles and to Ophiuchus with pink circles. We also show the 1:1 and $\pm$1 SpT relationships with the black solid and dashed lines, respectively.  A small offset in the SpT measurement from the literature has been applied to all sources for clarity.}
    \label{fig:spt_comparison}
\end{figure}

We derived the SpT and visual extinction ($A\mathrm{_V}$) simultaneously by comparing the X-Shooter spectra with spectral templates. We followed a similar methodology to that presented in \citet{manara15}. We compared the VIS and NIR arms target spectra with a set of Class III nonaccreting young objects with SpT K5-M9.5 observed with X-Shooter \citep{manara13temp,manara17temp}, which serve as zero-extinction templates ($A\mathrm{_V}$$<$0.2 mag). Both the target and template spectra are normalized at 1.05 $\mu$m and resampled to 3 nm wide median windows. We excluded the $K$-band from the fitting process to avoid any potential contribution from the disk. Instead, we do not allow the spectral template to have a $K$-band flux 10\% greater than that of the object spectrum. In the fitting process, we masked the regions that can be contaminated by telluric absorption and the region around H$\alpha$. We compare each object with all the spectral templates reddened by $A_\mathrm{V}$=0--24 mag with a 0.2 mag step. The goodness of fit of each comparison is evaluated using the reduced $\chi^2$ minimization:

\begin{equation}
\label{eqn:eq_chi2}
    \chi^2=\frac{1}{N-m} \sum_{i=1}^{N} \frac{ (O_i-T_i)^2} {\sigma^2 }
,\end{equation}

where $O$ is the object spectrum, $T$ the template spectrum, $\sigma$ is the noise of the observed spectrum, $N$ the number of data points, and $m$ the number of fitted parameters ($m$=2). To each object, we associate the SpT and $A\mathrm{_V}$ of the fit that minimized the $\chi^2$. However, in the case of CRBR 2317.5-172, the template that minimizes the $\chi^2$ is M9.5, which is significantly later than previous estimations (M5-M6, see Table~\ref{tab:tab_lit_params}). Closer inspection reveals that the M9.5 template reproduces well the $H$-band of the target spectrum but fails to reproduce the end of the $J$-band (1.2-1.35 $\mu$m), where the target presents a flatter spectrum that is more representative of earlier SpTs \citep{cushing05}. The $J$-band is better reproduced by SpTs earlier than M6.5. While the M6.5 template reproduces fairly well the $H$ band, earlier SpTs have much flatter $H$ bands. CRBR 2317.5-1729 is therefore classified as M6.5 with $A\mathrm{_V}$=17 mag. In Fig.~\ref{fig:spt_comparison}, we show the comparison of the derived SpTs with literature measurements (see Table~\ref{tab:tab_lit_params} for a compilation of the SpT measurements in the literature and the new measurements). We observe that both measurements are in very good agreement, with a root mean squared error of 0.58.

We also estimate the SpT using spectral indices defined in the optical and NIR ranges. In the optical, we use the indices selected in \citet{manara13temp} coming from \citet{riddick07}, without the c81 and PC3 indices which we found to provide very discrepant values in our SpT range of interest (M4-L0). These indices are calibrated up to $\sim$M8. In the NIR, we use the methodology presented in \citet{almendros22}, that includes six different spectral indices. All these spectral indices are affected by extinction, so before the derivation of the indices we de-redden the spectra using the extinction obtained in the comparison with the spectral templates. We observe that the VIS indices are within 1 subSpT up to M8, where the method stops working. The NIR indices, on the other hand, do a worse job at M4-M5 generally overestimating the value when compared to those obtained through a comparison with the templates. At later SpTs, the NIR SpT indices perform within 1 subSpT. Seeing that the SpT measurement coming from the comparison with spectral templates is in agreement with the SpT coming from spectral indices, we used the former as the SpT measurement for all of our sources.

\begin{table*}
    \caption{SpT, extinction, physical, and accretion properties of the sample presented in this work.}
    \begin{center}
        \begin{tabular}{l c c c c c c c c c}
            \hline\hline
            Name & Region & SpT & $A\mathrm{_V}$ & $T\mathrm{_{eff}}$ & $L\mathrm{_{*}}$ & log$L\mathrm{_{acc}}$ & $M\mathrm{_{*}}$ & log$M\mathrm{_{acc}}$ & Acc?$^a$ \\ 
             &  &  & [mag] & [K] & $L\mathrm{_{\odot}}$ & $L\mathrm{_{\odot}}$ & $M\mathrm{_{\odot}}$ & $M\mathrm{_{\odot}}$yr$^{-1}$ &  \\ 
            \hline
            CHSM 12653 & Cha-I & M7.5 & 1.0 & 2720 & 0.007 & -4.7 & 0.032 & -11.02 & N \\ 
            2MASS J11084952-7638443 & Cha-I & M9 & 1.4 & 2570 & 0.003 & -5.1 & 0.023 & -11.49 & N \\ 
            2MASS J11112249-7745427 & Cha-I & M9 & 0.0 & 2570 & 0.003 & -5.6 & 0.022 & -11.91 & N \\ 
            2MASS J11114533-7636505 & Cha-I & M8 & 0.2 & 2670 & 0.003 & -5.8 & 0.031 & -12.27 & Y \\ 
            2MASS J11122250-7714512 & Cha-I & M9 & 0.6 & 2570 & 0.003 & -5.8 & 0.023 & -12.19 & N \\ 
            2MASS J16031329-2112569 & Upper Sco & M4.5 & 0.6 & 3085 & 0.04 & -4.0 & 0.132 & -10.71 & Y \\ 
            2MASS J16052661-1957050 & Upper Sco & M4.5 & 0.4 & 3085 & 0.062 & -3.7 & 0.15 & -10.33 & Y \\ 
            2MASS J16053215-1933159 & Upper Sco & M4.5 & 0.8 & 3085 & 0.027 & -3.1 & 0.121 & -9.85 & N \\ 
            2MASS J16060061-1957114 & Upper Sco & M4 & 0.8 & 3190 & 0.093 & -3.7 & 0.195 & -10.39 & Y \\ 
            PGZ2001 J160702.1-201938 & Upper Sco & M5 & 0.6 & 2980 & 0.033 & -3.8 & 0.094 & -10.31 & N \\ 
            2MASS J16083455-2211559 & Upper Sco & M4.5 & 0.4 & 3085 & 0.024 & -4.4 & 0.12 & -11.12 & Y \\ 
            2MASS J16101888-2502325 & Upper Sco & M4.5 & 0.6 & 3085 & 0.047 & -4.0 & 0.142 & -10.64 & Y \\ 
            2MASS J16102819-1910444 & Upper Sco & M4.5 & 1.0 & 3085 & 0.027 & -4.0 & 0.123 & -10.71 & Y \\ 
            2MASS J16145928-2459308 & Upper Sco & M4 & 0.6 & 3190 & 0.067 & -4.0 & 0.192 & -10.7 & Y \\ 
            2MASS J16151239-2420091 & Upper Sco & M4.5 & 0.6 & 3085 & 0.014 & -4.4 & 0.111 & -11.27 & Y \\ 
            2MASS J16181618-2619080 & Upper Sco & M4 & 2.0 & 3190 & 0.055 & -3.9 & 0.189 & -10.69 & Y \\ 
            CRBR 2317.5-1729 & Ophiuchus & M6.5 & 17.0 & 2815 & 0.099 & -3.3 & 0.087 & <-9.48 & ? \\ 
            CRBR 2322.3-1143 & Ophiuchus & M5 & 10.8 & 2980 & 0.012 & -4.5 & 0.079 & <-11.22 & ? \\ 
            ISO-Oph042 & Ophiuchus & M5 & 6.6 & 2980 & 0.043 & -3.2 & 0.098 & -9.65 & N \\ 
            GY92 80 & Ophiuchus & M5.5 & 6.3 & 2920 & 0.022 & -4.4 & 0.07 & <-10.88 & ? \\ 
            GY92 90 & Ophiuchus & M9 & 8.5 & 2570 & 0.008 & -4.6 & 0.02 & <-10.71 & ? \\ 
            SONYC RhoOph-6 & Ophiuchus & M9.5 & 8.4 & 2520 & 0.002 & -4.4 & 0.02 & <-10.64 & ? \\ 
            GY92 264 & Ophiuchus & M7.5 & 1.0 & 2720 & 0.012 & -4.1 & 0.033 & -10.33 & N \\ 
            CFHTWIR-Oph 77 & Ophiuchus & M8 & 11.8 & 2670 & 0.001 & -3.9 & 0.047 & <-10.69 & ? \\ 
            BKLT J162736-245134 & Ophiuchus & M9.5 & 5.8 & 2520 & 0.002 & -4.3 & 0.023 & <-10.72 & ? \\ 
            BKLT J162848-242631 & Ophiuchus & M6.5 & 2.2 & 2815 & 0.01 & -4.9 & 0.045 & -11.36 & Y \\ 
            \hline 
        \end{tabular}
        \\$^a$Accretion measurement may have contribution from the chromospheric activity (see Sect~\ref{analysis_accreting}), possible non-accretors are reported with “Y”, whereas upper limit estimates are reported with “?”.
    \end{center}
    \label{tab:tab_results}
\end{table*}

\subsection{Stellar properties}
\label{analysis_prop}

The luminosity of the spectral templates has been carefully estimated in the literature \citep{manara13temp,manara17temp}. Therefore, we estimated the luminosity of our sources from that of the spectral templates: $L_*=d_*^2/d_{temp}^2F_N\cdot  L_{temp}$, where $d_*$ and $d_{temp}$ are the distances to the target source and the spectral template source, $F_N$ is the flux scaling factor between the template and the de-reddened flux calibrated target spectrum, and $L_{temp}$ is the luminosity of the spectral template. For each individual object we take the distances from \textit{Gaia} DR3 if they are available and parallax/$\sigma_{parallax}$ $>$ 10, otherwise we take the mean distance to the region. We derive the mean distance to each region using $Gaia$ DR3 parallaxes and the literature census of members of these regions: for Upper Scorpius and Ophiuchus, we take the membership census of \citet{luhman22} and for Cha-I the membership census of \citet{esplin17}. We derived the distance to each cluster and its uncertainty from the mean and standard deviation of the inverse of the parallax of the sources with parallax/$\sigma_{parallax}>$10.  We find distances of 145$\pm$9 pc, 138$\pm$5 pc, and 190$\pm$13 pc for Upper Scorpius, Ophiuchus, and Cha-I, respectively. The error in luminosity is calculated from error propagation of the distance errors and an error in the templates luminosity of 0.2 dex \citep{manara13temp}. 

We calculate the $T_{\mathrm{eff}}$ of all the targets directly from their SpT. We use the $T_{\mathrm{eff}}$-SpT scale of \citet{herczeg14} for the objects with SpT earlier than M8 in order to be consistent with \citet{manara23}, which is the main catalog that we compare our measurements to (see Sect.~\ref{results_literature}). At later SpTs, we use the scale of \citet{luhman03a}. We did this since we observe that the scale of \citet{herczeg14} undergoes a flattening of the relationship at late SpTs that is not found in any other scale \citep{pecaut13,mentuch08,luhman03a,faherty16}. The $T_{\mathrm{eff}}$-SpT scale found in \citet{faherty16}, derived using members of nearby young moving groups (ages 5-130 Myr) in the M7-L8 SpT range, hints that young objects follow the steepening seen in \citet{luhman03a} rather than that of \citet{herczeg14}. The error in $T_{\mathrm{eff}}$ is propagated from the SpT error. The SpT error is assumed to be the SpT step of the spectral templates at the SpT value.

\begin{figure}[hbt!]
    \centering
    \includegraphics[width=\textwidth/21*10]{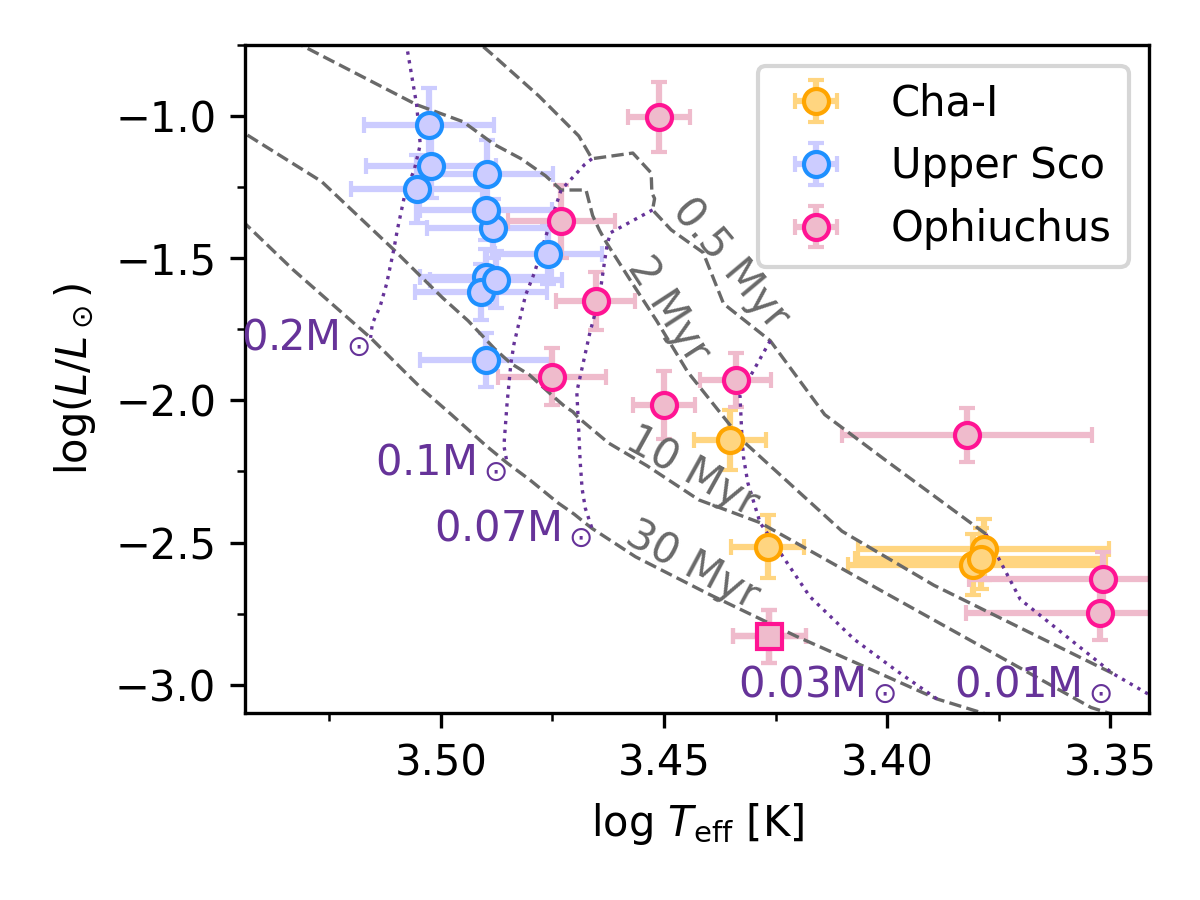}
    \caption{HR diagram of the sample presented in this work. The sources are color-coded following the same color-coding as in Fig.~\ref{fig:spt_comparison}. The isochrones (black dashed lines, shown for 0.5, 2, 10, and 300 Myr) and the lines of constant mass (dark blue dotted lines, shown for 0.04, 0.07, 0.1, and 0.2 $M_\odot$) shown are from \citet{baraffe15}. Source CFHTWIR-Oph 77 is shown with a square to highlight its possible subluminous nature.}
    \label{fig:hrd}
\end{figure}

In Fig.~\ref{fig:hrd}, we show the Hertzsrpung-Russell (HR) diagram for the entire sample together with the isochrones from \citet{baraffe15}. All the sources are located in the HR diagram within the typical spread observed in previous studies of the same regions \citep{manara15,manara17cha,manara20}. Object CFHTWIR-Oph 77, Ophiuchus member, is located below the 30 Myr isochrone, which could indicate that this object is subluminous. In the literature this object was classified with M9.75 \citep{alvesoliveira12}, whereas we have classified it as M8. However, the X-Shooter spectrum of this object does not present good signal-to-noise ratio (S/N) even in the NIR arm (see Table~\ref{tab:tab_obs}). Using a SpT of M9.75 would place this object over the 2 Myr isochrone in the HR diagram. For consistency, we maintain our spectral classification and consider the object to be subluminous.

In order to estimate the masses, we first interpolated the \citet{baraffe15} isochrones to a finer grid of 1000$\times$1000 points in age and mass. Then, for each target we perform a monte carlo simulation with 1000 iterations. In each iteration we perform a random sampling of the target's $T\mathrm{_{eff}}$ and $L_*$ in accordance with the respective uncertainties and the selected age and mass will be that of the closest point to the interpolated isochrones. The mass and errors of each target are derived from the median and 25/75 percentiles, respectively.

\subsection{Accretion luminosity}
\label{analysis_lines}

We use accretion-related emission lines as the main proxy for the measurement of the accretion luminosity of the targets, which have been shown to provide consistent measurements of accretion, specially when considering multiple lines \citep{rigliaco12,alcala14}. Only the Upper Scorpius targets and one of the Ophiuchus targets present enough signal in the UVB arm to inspect the accretion rates through the Balmer continuum excess emission and the Balmer jump (see Sect.~\ref{analysis_uv}).

We first visually identify the emission lines that are detected in each of the targets, focusing on 10 lines that are known to be bright and have a good correlation with $L\mathrm{_{acc}}$: CaK, H$\delta$, H$\gamma$, H$\beta$, He$\lambda$587nm, H$\alpha$, He$\lambda$667nm, Pa$\gamma$, Pa$\beta$, and Br$\gamma$. The line fluxes are estimated on the flux-calibrated de-reddened spectra using an automatic method that estimates the continuum and the line extent simultaneously \citep{manara15}. Since this process is automatic, we visually checked that the extent of all the lines is correctly estimated and we corrected them if needed. The continuum is then subtracted from the spectrum. For each line detected in emission we calculate the integrated flux over the line extent. The error in the line flux is estimated from the propagation of 1$\sigma$ of the estimated continuum. For the sources with no detected emission in any line we derive the upper limits of the accretion luminosity as 3 $\times$ $F\mathrm{_{noise}} \times \Delta \lambda$, where $F\mathrm{_{noise}}$ is the rms flux-noise in the region of the line and $\Delta\lambda$ is the expected average line width, assumed to be 0.2 nm. The accretion luminosity measured using different lines agree well with each other (typically within 0.2 dex).

Using the known distance to the targets we convert the line flux to line luminosity. Lastly, we compute the accretion luminosity using the $L_{\mathrm{line}}$--$L_{\mathrm{acc}}$ relationships of \citet{alcala17}. The accretion luminosity of each source is determined as the mean of the $L\mathrm{_{acc}}$ of all the detected lines, and its uncertainty from the standard deviation of the $L\mathrm{_{acc}}$ values. The accretion luminosity of each source is converted into mass accretion rates ($\dot{M}\mathrm{_{acc}}$) using the following relation \citep{gullbring98}:

\begin{equation}
    \dot{M}\mathrm{_{acc}} = \frac{L\mathrm{_{acc}}R_*}{GM_*(1-\frac{R_*}{R\mathrm{_{in}}})}
,\end{equation}

where $R\mathrm{_{in}}$ denotes the truncation radius of the disk (assumed to be 5~$R\mathrm{_*}$), $G$ is the gravitational constant, $M\mathrm{_*}$ is the stellar mass, and $R\mathrm{_*}$ is the stellar radius. The latter is estimated from the $T\mathrm{_{eff}}$ and luminosity assuming black-body radiation: $L_*=4\pi\sigma\mathrm{_B} R_*^2T\mathrm{_{eff}}^4$.

Eight of the targets in Upper Scorpius analyzed here overlap with the \citet{fang23} sample. The accretion luminosities and accretion rates derived in \citet{fang23} agree within 0.2 dex with our measurements.

\subsection{UV excess}
\label{analysis_uv}

Additionally, for the targets that have good S/N in the UVB arm (all Upper Scorpius targets and GY92 264), we also derive $L_{\mathrm{acc}}$ from modeling of the UV continuum following the methodology presented in \citet{manara13acc}. This methodology performs a simultaneous fitting of a photospheric template (the same ones we used in Sect.~\ref{analysis_templates}) and a slab model that aims to reproduce the excess emission caused by the accretion \citep[see][]{manara13acc}. This kind of modeling of the accretion emission is able to reproduce most of the accretion observables and has been used extensively \citep[e.g.,][]{herczeg08,rigliaco12,manara13acc,alcala17,manara16,manara17cha,rugel18,venuti19}.

These measurements are consistent with the accretion rate measurements using the emission lines made in Sect.~\ref{analysis_lines}, with some exceptions: Several sources present negligible UV excess (see Sect.~\ref{analysis_accreting}) and their UV modeling provides significantly lower accretion luminosities than those obtained from the emission lines. We decided to use the accretion luminosity measurements coming from emission lines as the measurement for accretion luminosity for all the sources for consistency within our sample.

\subsection{Accreting objects}
\label{analysis_accreting}

\begin{figure*}[hbt!]
    \centering
    \includegraphics[width=\textwidth]{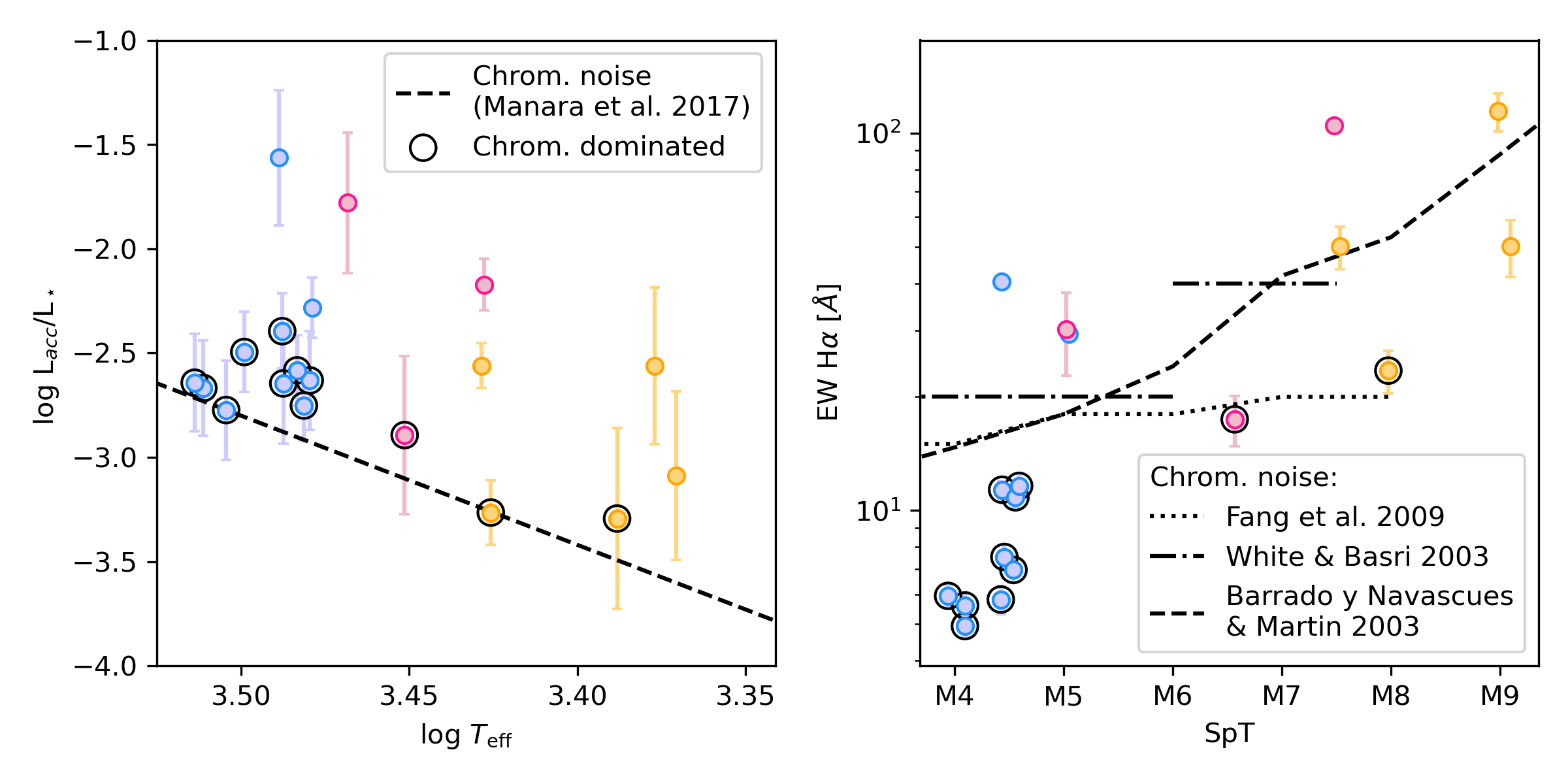}
    \caption{\textit{} Logarithm of the ratio between the accretion luminosity and the stellar luminosity as a function of log$T\mathrm{_{eff}}$ (left panel). The blak dashed line represents the boundary below which the impact of chromospheric emission on the observed accretion signatures becomes dominant. The symbols follow the same color-coding as in Fig.~\ref{fig:spt_comparison}. EW of the H$\alpha$ emission line as a function of SpT (right panel). We show the chromospheric-dominated boundary from \citet{fang09} with a black dotted line and the one from \citet{white03} with a dash-dotted line. We also show the saturation limit from \citet{barrado03} with a dashed line. In both panels we highlight the sources that may have a significant contribution in the measured accretion properties from the chromospheric activity with large empty black circles. A small offset in log$T\mathrm{_{eff}}$ (left panel) and SpT (right panel) has been applied to all sources for clarity.}
    \label{fig:lacclstar}
\end{figure*}

An important caveat that needs to be addressed is that the emission lines may not be tracing accretion of mass from the protoplanetary disk but chrompospheric activity of the central object. Pre-main sequence objects also possess high chromospheric activity and emission from the chromosphere contributes to the emission in both the UV continuum and the emission lines. Chromospheric activity can even dominate the emission in objects with low accretion rates. \citet{manara13temp} and \citet{manara17temp} studied the intensity of chromospheric emission in non-accreting young objects like the ones we used for spectral typing in Sect.~\ref{xshoo_analysis}. In the left panel of Fig.~\ref{fig:lacclstar}, we show the log($L_{\mathrm{acc}}/L_{\mathrm{*}}$) ratios of our sample with detected accretion rate. We also show the approximate emission level below which the contribution from stellar chromospheric activity may become predominant as measured in \citet[shown with a black dashed line]{manara17temp}. We observe that only half of the targets have $L\mathrm{_{acc}}/L\mathrm{_*}$ larger than the $L\mathrm{_{acc}}/L_*$ chromospheric boundary by more than 1$\sigma$.

Another commonly used tracer that can discriminate accretion from chromospheric dominated emission is the EW of the H$\alpha$ emission line \citep{white03,barrado03,fang09}. In the right panel of Fig.~\ref{fig:lacclstar}, we show the H$\alpha$ EW as a function of SpT for our sample. We also show the criteria for distinguishing classical and weak line T Tauri stars of \citet{fang09}, \citet{white03} and \citet{barrado03}, that is sources with accretion-dominated emission from sources with chromospheric-dominated emission. We observe that all the Upper Scorpius sources except two (2MASS J16053215 and PGZ160702) lie well below the chromospheric boundary, and one Ophiuchus (BKLT J162848) and another Cha-I (2MASS J11114533) sources are below or very close to the \citet{fang09} chromospheric noise boundary. There is one Cha-I source (2MASS J11122250) that has negligible continuum at the location of the H$\alpha$ line, therefore its EW measurement is highly sensitive to the continuum that is subtracted to the spectrum (by orders of magnitude). Additionally, the criterion of \citet{fang09} is only defined up to M8 and that of \citet{white03} up to M7.5. Above that SpT, we can use the \citet{barrado03} saturation criterion, which finds another Cha-I source to be be consistent with chromospheric noise (2MASS J11112249).

The width of the H$\alpha$ emission line is also a good tracer of accretion \citep{muzerolle98lines} and is typically inspected using the full width of the line at 10\% of the peak intensity. If the width of the H$\alpha$ line is larger than 200 km/s, then the emission is most probably associated with accretion \citep{jayawardhana03}. The Cha-I source with SpT$>$M8 and very uncertain H$\alpha$ EW measurement (2MASS J11122250) presents a H$\alpha$ full width smaller than 200 km/s. All of the sources with SpT$\leq$M8 located below the H$\alpha$ EW chromospheric boundary present a full width of the H$\alpha$ line at 10\% of the peak intensity smaller than 200 km/s. Source 2MASS J11112249, which appeared below the \citet{barrado03} EW H$\alpha$ criterion, has an H$\alpha$ emission line width of $\sim$200 km/s. This source was also above the log($L_{\mathrm{acc}}/L_{\mathrm{*}}$) chromospheric noise criteria by slightly more than 1 $\sigma$. Therefore, given that the source meets two of the criteria, we consider the source to be an accretor.

Lastly, we use the shape of the HeI $\lambda$1083 line to discriminate between accretion and chromospheric activity \citep{thanathibodee22}. If a target presents blueshifted or redshifted features in this line, they can be associated to accretion, in contrast with the chromospheric feature that is located at the line's center \citep{thanathibodee22,edwards06}. A visual evaluation of the HeI $\lambda$1083 line of the sources with detected accretion rates leads to the same conclusion found in the study of the H$\alpha$ emission line width.

Here, we briefly discuss some objects where the accretion origin of their emission lines has been assessed in the literature. Object ISO-Oph042 was classified in \citet{mohanty05} as a non-accretor based on the H$\alpha$ 10\% width. We find this object to present redshifted features in the HeI $\lambda$1083 line and to meet all the accretion diagnostics. Object PGZ2001 J160702.1-201938 was classified in \citet{fang23} as a non-accretor based on the EW and 10\% full-width of the H$\alpha$ line. However, we find this object to be above the chromospheric boundary in all accretion diagnostics and to present some redshifted features in the HeI $\lambda$1083 line. The rest of the targets from Upper Scorpius presented here agree with the \citet{fang23} accretion classification. \citet{thanathibodee22} found source 2MASS J16060061-1957114 to be consistent with chromospheric activity based on the HeI $\lambda$1083 line shape, as we also did.

Overall,  we consider two Upper Scorpius sources (2MASS J16053215 and PGZ J160702.1), three Cha-I sources (2MASS J11084952, 2MASS J11112249, and CHSM 12653) and all but BKLT J162848 from Ophiuchus as accretors (see Table~\ref{tab:tab_results}). The emission of the rest of the targets may have a significant contribution from chromospheric activity, so we consider them as possible accretors and treat their accretion measurement as an upper limit (marked with large black empty circles in Fig.~\ref{fig:lacclstar}).

\subsection{Literature data}
\label{results_literature}

Here, we compile the available measurements of the accretion rate in the four star-forming regions studied in this work: Ophiuchus, Lupus, Cha-I, and Upper Scorpius. \citet{manara23} performed a compilation of the available accretion luminosity ($L\mathrm{_{acc}}$) and protoplanetary dust mass ($M\mathrm{_{dust}}$) measurements of sources in nearby star-forming regions ($<$300 pc). The stellar and accretion parameters compiled in that work came mainly from surveys carried out with X-Shooter. The authors re-scaled the stellar luminosities to the new \textit{Gaia} distances and the $T\mathrm{_{eff}}$ was obtained from the SpT using the scale from \citet{herczeg14}. Stellar masses were derived using the \citet{baraffe15} models for $T\mathrm{_{eff}}<$3900 K and of \citet{feiden16} for hotter stars. In a few cases where the stellar properties were outside the range of validity of the models, the models from \citet{siess00}  were used. The authors also performed a compilation of protoplanetary disk dust content measurements, which come from many different ALMA surveys, and homogenized the measurements following \citet{ansdell16}. We use this compilation as the primary source for data from the literature. Regarding the Ophiuchus star-forming region, we only use the data of known members of the L1688 cloud, which is the youngest in the region \citep{wilking08,esplin20}.

The main change of the stellar and accretion measurements of \citet{manara23} from previous accretion rate surveys using X-Shooter is the SpT-$T\mathrm{_{eff}}$ scale used. Previous X-Shooter surveys of the same star-forming regions used the \citet{luhman03a} scale at M-types and the \citet{kenyon95} scale for the K-type stars \citep{manara15,manara17cha,alcala17,manara20,testi22}, while \citet{manara23} used the \citet{herczeg14} scale for the entire SpT range. At SpTs earlier than K8 ($\geq$0.65 $M_\odot$), the \citet{kenyon95} scale systematically produces higher $T\mathrm{_{eff}}$, and the difference between the two scales is larger than 100 K at SpTs earlier than K6. The same happens at SpTs M4-M8 ($\sim$0.03-0.2 $M_\odot$), where the two scales have a typical difference of 100 K. This is also the SpT range where a change in $T\mathrm{_{eff}}$ corresponds to the greatest change in mass in the \citet{baraffe15} isochrones ($\Delta M_* \approx$0.2 dex for a variation in $T\mathrm{_{eff}}$ of 100 K). In both SpT ranges, the difference between the $T\mathrm{_{eff}}$ scales translates to different masses for the same SpT (lower masses using the \citealt{herczeg14} scale), which affects the mass accretion rate measurements and the classification of objects as stellar or sub-stellar as well. Both scales agree well in the SpT range K8-M3 which roughly translates to masses 0.3-0.65 $M_\odot$. Overall, the $M_*-\dot{M}\mathrm{_{acc}}$ diagram will shift toward lower masses and higher accretion rates when using the \citet{herczeg14} scale in the two SpT ranges, where the scales differ the most (as mentioned above) and the relationship will conserve its shape in the 0.3-0.65 $M_\odot$ mass range.

Recently, \citet{fang23} studied a sample of Upper Scorpius members using Keck/HIRES. The accretion properties were derived from the line luminosities of the emission lines present in the optical wavelength range (480-980 nm) using the $L\mathrm{_{line}}-L\mathrm{_{acc}}$ relationships of \citet{fang18}. \citet{fang18} scaled the \citet{alcala17} $L\mathrm{_{line}}-L\mathrm{_{acc}}$ relationships to the \textit{Gaia} DR2 distances. \citet{fang23} derived the $T\mathrm{_{eff}}$ directly from the SpT using the \citet{fang17} scale, which is combination of that from \citet{pecaut13} for stars earlier than M4 and from \citet{herczeg14} for stars later than M4. The masses were derived using the \citet{feiden16} evolutionary tracks. We used the combined \citet{manara23} and \citet{fang23} sample as a literature source for accretion rate measurements in Upper Scorpius. For the sources that appear in both catalogs we take the accretion measurements from \citet{manara23} that come from the modeling of the UV emission continuum.

\citet{majidi23} recently measured the accretion rates of several new members of the Lupus star-forming region using X-Shooter. They obtained the physical parameters of the sources using ROTFIT \citep{frasca17}. The accretion luminosity was estimated from the luminosity of several emission lines and the \citet{alcala17} $L\mathrm{_{line}}-L\mathrm{_{acc}}$ relationships. Since these sources are new members of the Lupus star-forming region, they have no available measurement of the mass of their protoplanetary disks.

We evaluated the membership of all the sources we have used to its corresponding star-forming region based on their proximity to the known members of the region and their kinematics. In Sect.~\ref{lupus_membership}, we provide a detailed report of this analysis. Overall, we found a number of Lupus sources to be possible non-members of the region, but rather of one of the older associations of the Scorpius-Centaurus complex. We consider these sources as possible non-members of the Lupus star-forming region and discuss throughout the paper how the rejection of these sources as members of Lupus would affect the results.

The entire sample of sources with accretion rate measurement and/or protoplanetary dust content estimate from the literature combined with the new accretion rate estimates using the X-Shooter spectra presented here will be used to study the relationship between the stellar properties, the mass accretion rate, and the protoplanetary dust content. In order to do so, all the sources needs to be analyzed using the same methodology. We therefore re-calculate the accretion rate and stellar masses of the \citet{fang23} and \citet{majidi23} sources following the same methodology we have used for our X-Shooter sample presented in this Section. Additionally, Lupus member 2MASS J16085953-3856275 has an SpT of M8.5 \citep{alcala14}. To be consistent with the methodology used in this work, we re-calculated its accretion rate and stellar mass using the \citet{luhman03a} SpT-$T\mathrm{_{eff}}$ relationship to estimate its $T\mathrm{_{eff}}$.

\begin{figure*}[hbt!]
    \centering
    \includegraphics[width=\textwidth]{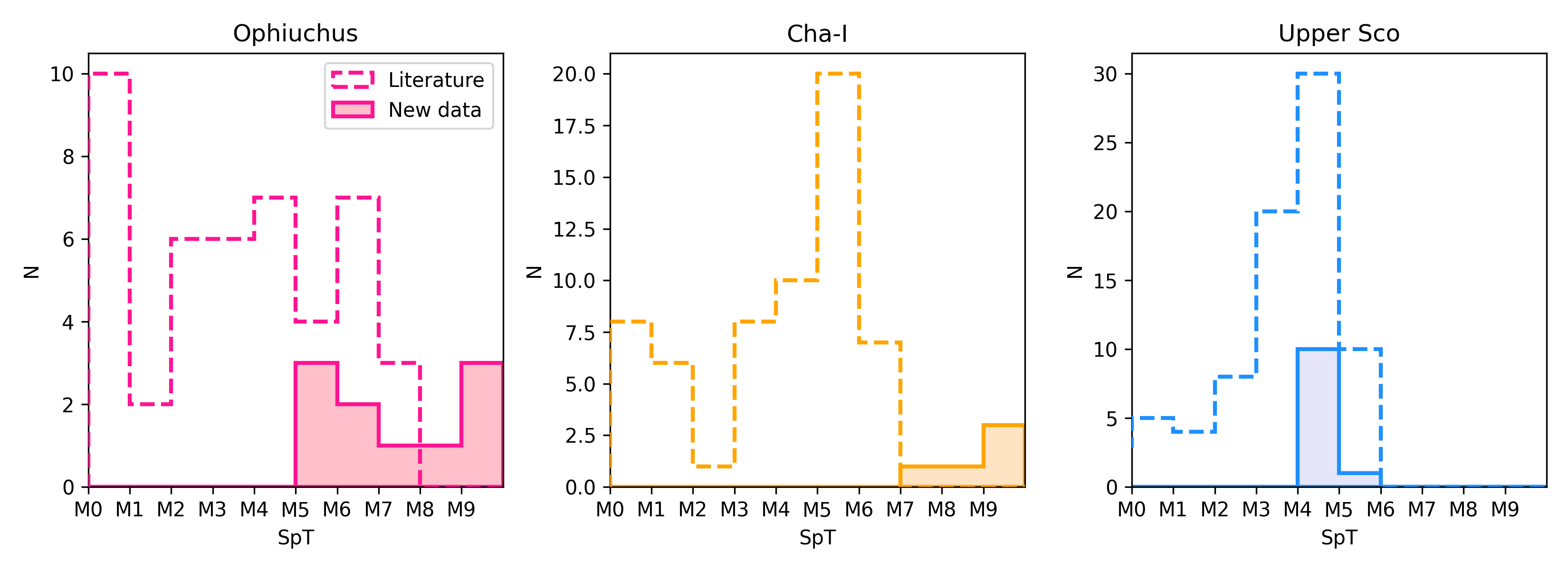}
    \caption{SpT histogram of sources with accretion measurement in the literature (dashed lines, see Sect.~\ref{results_literature}) and the new data (solid lines) in Cha-I (left panel), Upper Scorpius (middle panel), and Ophiuchus (right panel). Each region follows the same color-coding as in Fig.~\ref{fig:spt_comparison}.}
    \label{fig:spt_hist}
\end{figure*}

We also evaluated whether any of the sources in the literature are subluminous. We considered all subluminous sources that have been flagged as such in \citet{manara23} or if they are located below the 30 Myr isochrone (using \citealt{baraffe15} isochrones). Apart from the sources defined as subluminous in the literature, we also found one source from Ophiuchus (SSTc2d J162656.3-244120), two sources from Upper Scorpius (SSTc2d J162459.8-245601, SSTc2d J162637.1-241560), and one source from Lupus (2MASS J16085373-3914367) to be subluminous.

We also evaluated whether any of the sources in the literature are subluminous. We considered all subluminous sources that have been flagged as such in \citet{manara23}. We also found one source from Ophiuchus (SSTc2d J162656.3-244120), two sources from Upper Scorpius (SSTc2d J162459.8-245601, SSTc2d J162637.1-241560), and one source from Lupus (2MASS J16085373-3914367) to be located below the 30 Myr isochrone (using \citealt{baraffe15} isochrones), and considered them as subluminous as well.

The combined sample of sources with accretion rate measurement in the literature of the Ophiuchus, Lupus, Cha-I and Upper Scorpius star-forming regions represent $\approx$50\%, $>$90\%, $>$90\%, and $<$20\% of the entire census of sources with infrared excess in each region, respectively. While the census of Cha-I and Lupus members with accretion measurements is almost complete, the census in Ophiuchus and Upper Scorpius is highly incomplete, specially in the case of Upper Scorpius. As can be seen in Fig.~\ref{fig:spt_hist}, the new observations with X-Shooter increase significantly the amount of sources in the late spectral type bins with their accretion rate characterized in Cha-I, Ophiuchus, and Upper Scorpius.

\subsection{Relative ages}

This work is aimed at studying the evolution of protoplanetary disks. Therefore, we need to establish the relative age ordering of the four regions studied in this work. \citet{testi22} recently estimated the ages of the four regions from their HR diagrams and comparison to the \citet{baraffe15} evolutionary tracks. These authors found that the ordering in age of these regions would be (from youngest to oldest): Ophiuchus (1 Myr), Lupus (2 Myr), Cha-I (2.8 Myr), and Upper Scorpius (4.3 Myr). We performed the same analysis using the updated luminosities and $T\mathrm{_{eff}}$ and find the same ordering of the regions, although it is important to mention that Lupus and Cha-I have a very similar age. It is also worth noting that the quoted estimated age for Upper Scorpius is on the lower limit of recent age estimates of the region in the literature using different methodologies \citep[5-12 Myr,][]{ratzenbock23,luhman22,miret22}. Overall, we decided to follow the age ordering of \citet{testi22}.

\section{Results}
\label{xshoo_results}

In this section, we analyze the relationship between the accretion luminosity (mass accretion rate) and the stellar luminosity (mass), as well as the relationship of the protoplanetary disk dust mass with the stellar mass and the mass accretion rate. These relationships are built for all the sources with measured accretion properties in the Ophiuchus, Lupus, Cha-I, and Upper Scorpius star-forming regions.

\subsection{Relationship between luminosity and accretion luminosity}
\label{results_lstar_lacc}

\begin{figure*}[hbt!]
    \centering
    \includegraphics[width=\textwidth]{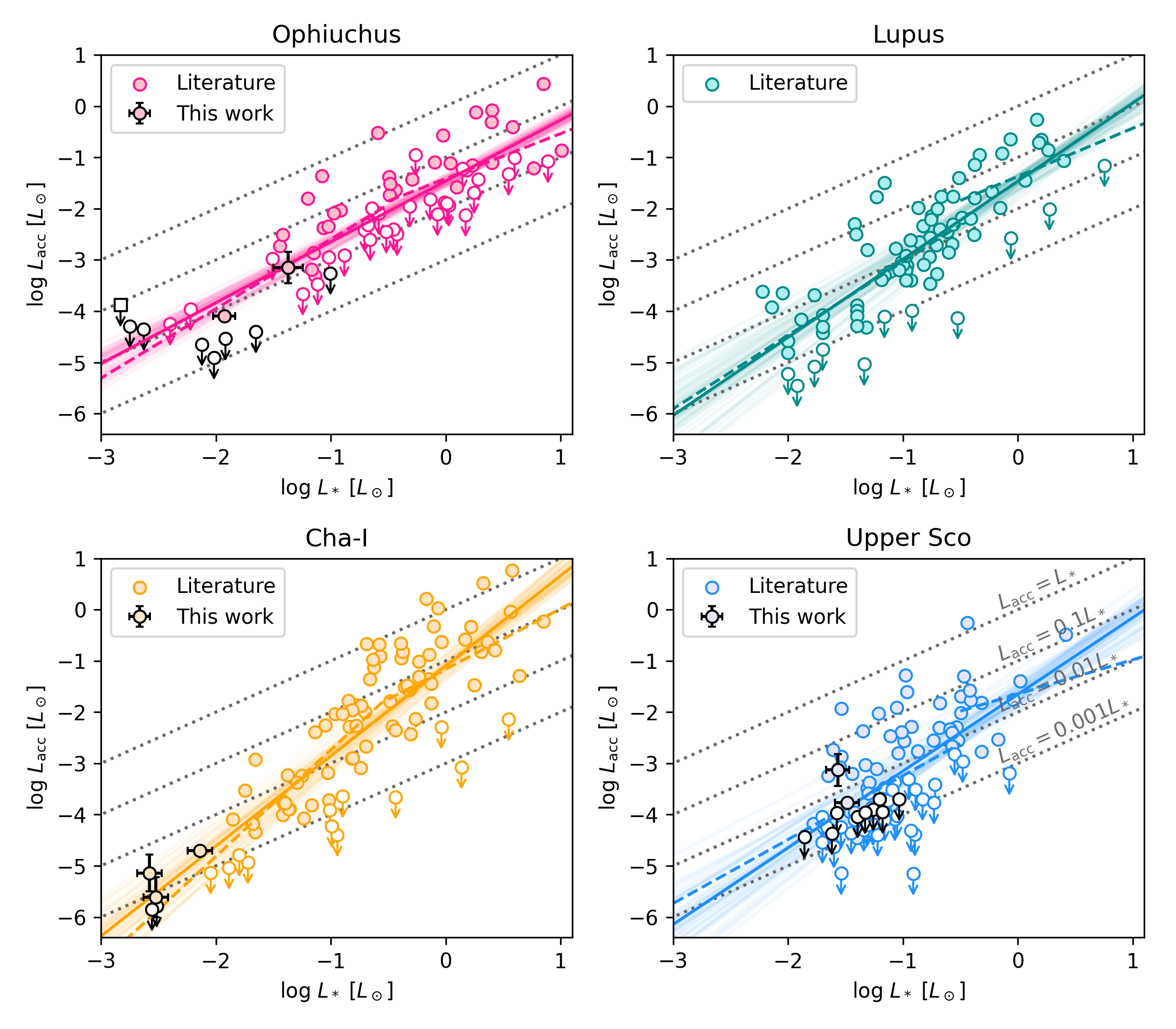}
    \caption{log$L_*-$log$L\mathrm{_{acc}}$ relationship for the four regions studied in this work: Ophiuchus (top left panel, pink), Lupus (top right panel, dark cyan), Cha-I (bottom left panel, orange), and Upper Scorpius (bottom right panel, blue). The data from the literature are shown in different colors, while the data presented in this work are shown with black circles. CFHTWIR-Oph 77 is represented in the top left panel with an empty black square to highlight its subluminous nature. We show the single and segmented power-law fits we have performed to the relationship in each region with solid and dashed lines, respectively. The light solid lines represent examples from the posterior distribution of the single power-law fit. The gray dotted lines mark the regions of constant $L\mathrm{_{acc}}/L_*$ ratio (labeled in the lower-right panel).}
    \label{fig:lstar_lacc}
\end{figure*}

In Fig.~\ref{fig:lstar_lacc}, we show the log$L_*-$log$L\mathrm{_{acc}}$ relation for the Ophiuchus, Lupus, Cha-I, and Upper Scorpius star-forming regions. In each panel, we present the data presented in this work in black  and  the data from the literature in different
colors (see Sect.~\ref{results_literature}; Ophiuchus, Cha-I, and Upper Scorpius follow the same color-coding of Fig.~\ref{fig:spt_hist}, while Lupus sources are shown with dark cyan circles). In these four regions we observe that the accretion luminosity increases with the luminosity of the central object. Higher luminosity sources also exhibit higher $L\mathrm{_{acc}} / L_*$ ratios ($L\mathrm{_{acc}}>$0.1$L_*$). The correlation between both parameters is therefore steeper than linear (i.e., slope larger than one of the linear correlation between the variables in log scale), in agreement with previous studies \citep[e.g.,][]{natta06,manara12}. The new observations presented in this work expand the $L\mathrm{_{acc}}$ measurements at the low-brightness edge of the Ophiuchus, Cha-I, and Upper Scorpius star-forming regions.

In order to quantify the relationship between $L_*$ and $L\mathrm{_{acc}}$, we performed a power-law fit using the maximum-likelihood Bayesian method developed by \citet{kelly07}\footnote{We used the available python implementation of the code: \url{https://linmix.readthedocs.io/en/latest/} and \url{https://github.com/jmeyers314/linmix}.}. The main advantage of this method is that it can handle upper limits data points, which dominate the available $L\mathrm{_{acc}}$ measurements in some regions. We performed the fit to the entire mass range spanned by all the observations in each region. Sub-luminous sources may be seen under a particular star-disk geometry (e.g., edge-on) and, as a result, their luminosity may not represent the total luminosity of the central source. Therefore, we excluded the subluminous sources in this or any of the fits performed in this work. In Fig.~\ref{fig:lstar_lacc}, we show the mean power-law fit in pink, dark cyan, orange, and blue solid lines for Ophiuchus, Lupus, Cha-I, and Upper Scorpius, respectively. The parameters of the single power-law fits performed are presented in Tab.~\ref{tab:tab_slopes}. The slopes found are in the range 1.2$-$1.8, confirming the steeper than linear correlation between log$L_*$ and log$L\mathrm{_{acc}}$. These slopes are also in the range of previous estimations \citep[e.g.,][]{natta14,alcala17}. The spread over the power-law fit is typically between 0.5 and 1 dex and is similar for the four regions studied.

There have been claims in the literature that the accretion relationship is rather better described with a double power law \citep{alcala17,manara17cha}. Therefore, we also modeled the $L_*-L\mathrm{_{acc}}$ relationship with a segmented power law broken at $L_*=10^{-0.5} L_\odot$. In Fig.~\ref{fig:lstar_lacc}, we show the broken power-law fits performed to each region with dashed lines. 

\begin{figure}[hbt!]
    \centering
    \includegraphics[width=\textwidth/21*10]{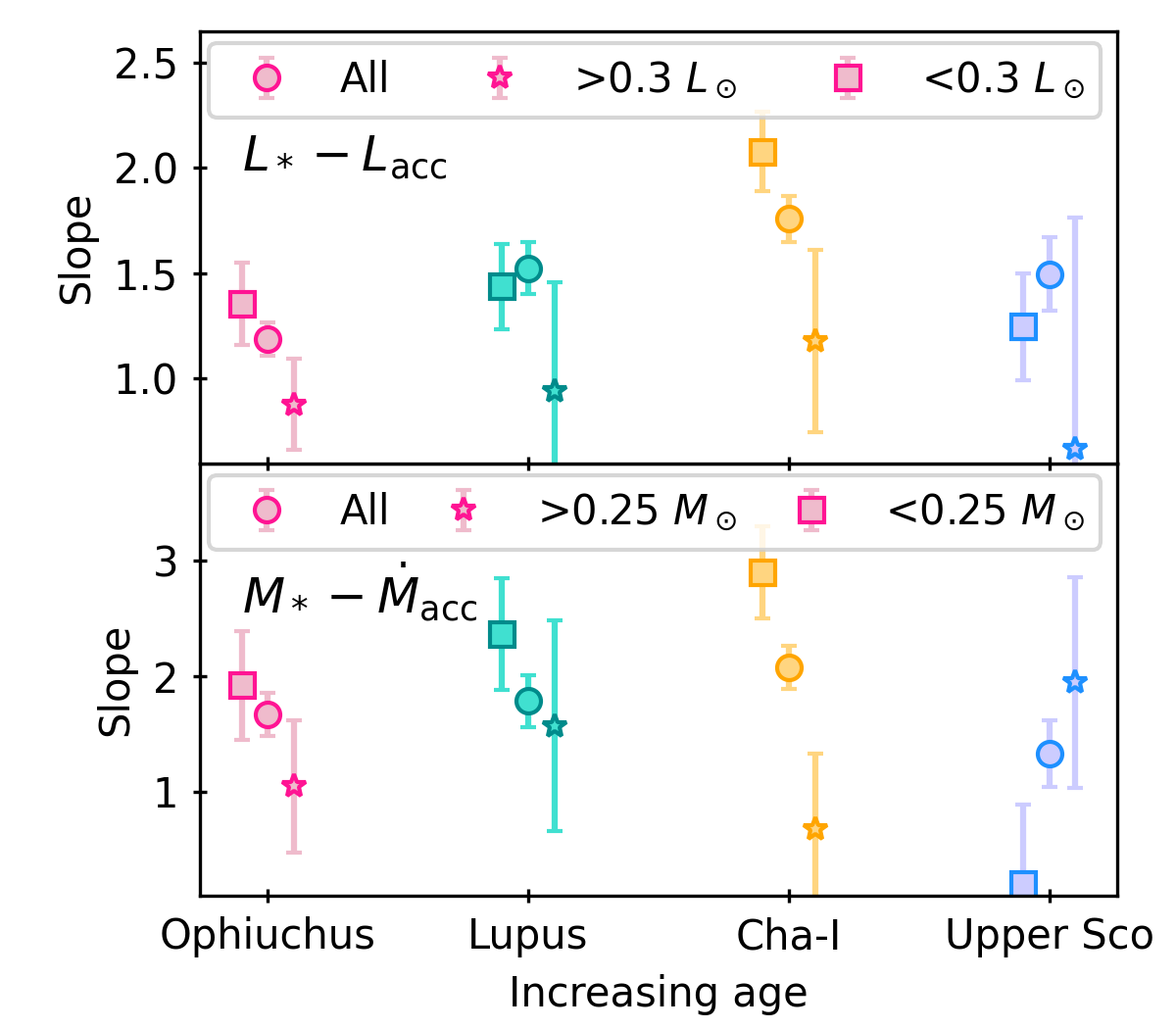}
    \caption{Slope of the single and segmented power-law fits to the log$L_*-$log$L\mathrm{_{acc}}$ (top panel) and log$M_*-$log$\dot{M}\mathrm{_{acc}}$ (bottom panel) relationships for the four regions studied in this work. The slopes of the single power-law fits are shown with circles, of the faint and low-mass samples with squares and the bright and high-mass samples with stars. A small offset in age in the slopes of each region was performed for visualization purposes. The symbols follow the same color-coding as in Fig.~\ref{fig:lstar_lacc}.}
    \label{fig:slope_age_mass}
\end{figure}

In the top panel of Fig.~\ref{fig:slope_age_mass}, we show the slopes of the single (represented with circles) and double power-law (represented with squares for the faint sample, and with stars for the bright sample) models of the $L_*-L\mathrm{_{acc}}$ relationship for each region. The slope of the single power-law model undergoes a steepening between Ophiuchus, Lupus and Cha-I (i.e., approximately between 1-3 Myr). The difference in the slope between these regions is larger than 1$\sigma$. Upper Scorpius does not continue the steepening trend found at younger ages and its slope is consistent with that of Lupus. The same result is found if we do not consider the possible non members of Lupus.

On the other hand, the slopes of the faint sample power-law fit in Ophiuchus, Lupus, and Cha-I present the same steepening trend seen in the single power-law fit in the 1-3 Myr age range, and the bright samples are consistent within 1$\sigma$ and with a slope$=1$ in this age range. The faint and bright samples of Upper Scorpius present slopes consistent with each other and with the single power-law slope. 

We used several metrics to test the statistical significance of the segmented power-law model over the single power-law fit in the four regions. We used the same metrics and criteria as in \citet{manara17cha}: the $R^2$, the Akaike information criterion (AIC), and the Bayesian information criterion (BIC; see e.g., \citealt{feigelson12}). We observe that the single power-law model is the preferred model in all the cases, although both models have very similar statistical significance. If we do not include the upper limits in the derivation of the metrics, the statistical preference of the single power-law model becomes even smaller. Since the single power-law model is simpler, we conclude that this model is a better representation of the $L_*-L\mathrm{_{acc}}$ relationship.

\subsection{Relationship between stellar mass and mass accretion rate}
\label{results_mstar_macc}

\begin{figure*}[hbt!]
    \centering
    \includegraphics[width=\textwidth]{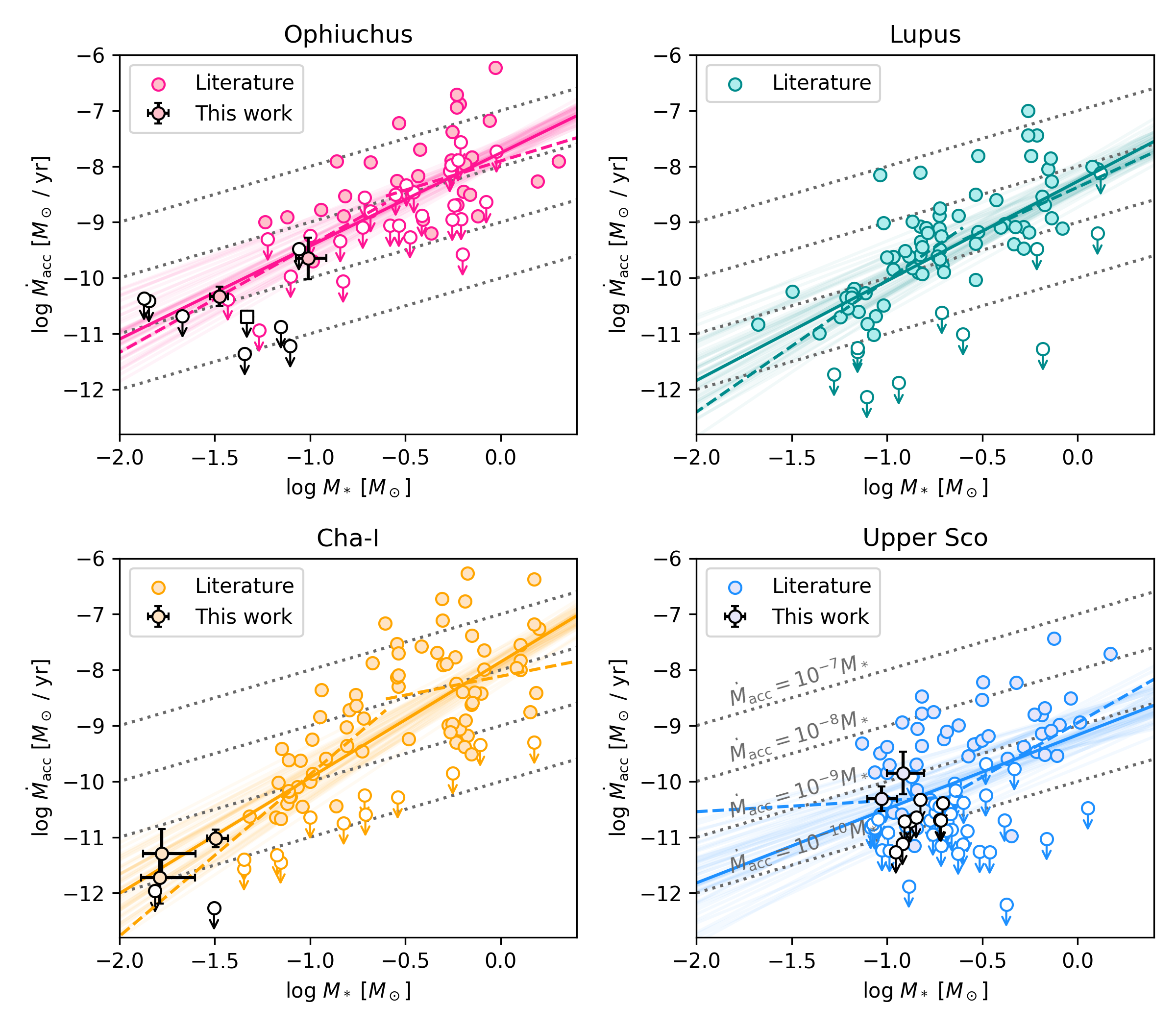}
    \caption{log$M_*-$log$\dot{M}\mathrm{_{acc}}$ relationship for the four regions studied in this work: Ophiuchus (top left panel), Lupus (top right panel), Cha-I (bottom left panel), and Upper Scorpius (bottom right panel). The symbols follow the same color-coding as in Fig.~\ref{fig:lstar_lacc}. The data from the literature are shown with different colors, while the data presented in this work are shown with black circles. CFHTWIR-Oph 77 is represented in the top left panel with an empty black square to highlight its subluminous nature. We also show the single and segmented power-law fits we have performed to the relationship in each region with solid and dashed lines, respectively. The light solid lines represent examples from the posterior distribution of the single power-law fit. The gray dotted lines mark the regions of constant $\dot{M}\mathrm{_{acc}}/M_*$ ratio (labeled in the lower-right panel).}
    \label{fig:mstar_macc}
\end{figure*}

In Fig.~\ref{fig:mstar_macc}, we show the log$M_*-$log$\dot{M}\mathrm{_{acc}}$ relation for the four regions studied in this work. We confirm that the mass accretion rate increases with the mass of the central object and that the correlation between the two parameters is steeper than linear, where more massive sources tend to have larger $\dot{M}\mathrm{_{acc}}/M_*$ ratios. The new observations presented here expand the $\dot{M}\mathrm{_{acc}}$ measurements at the low-mass tail of the Ophiuchus, Cha-I, and Upper Scorpius star-forming regions.

We performed a power-law fit to the log$M_*-$log$\dot{M}\mathrm{_{acc}}$ relationship of the four regions. The mean single power-law fit is represented in Fig.~\ref{fig:mstar_macc} with solid lines and the parameters of these fits are presented in Table~\ref{tab:tab_slopes}. The derived slopes are in agreement with the slopes found in the literature for the same regions \citep{alcala17,manara17cha,manara15,testi22,mulders17}. The spread over the power-law fit is typically between 0.5 and 1 dex, which is consistent among the four regions and with \citet{testi22}.

The correlation between $M_*$ and $\dot{M}\mathrm{_{acc}}$ undergoes an apparent steepening below $\sim$0.25 $M_\odot$ in Ophiuchus, Cha-I and Lupus. Such a behavior can imply that the relationship is better described by a double power law or by a single steeper slope following the upper envelope of accretion rates. In the latter case, as the $M_*$ decreases the range of $\dot{M}\mathrm{_{acc}}$ values (at a given mass) becomes smaller because the low accretors at this mass range would not present detectable $\dot{M}\mathrm{_{acc}}$ or disk. Therefore, we also model the log$M_*-$log$\dot{M}\mathrm{_{acc}}$ relationships with a double power law (dashed lines in Fig.~\ref{fig:mstar_macc}). In the bottom panel of Fig.~\ref{fig:slope_age_mass} we show the slopes of the single and double power-law models of this relationship. The single power-law slopes of this relationship do not present the same evolutionary sequence as the $L_*-L\mathrm{_{acc}}$ relationship: the slope of Cha-I is the steepest, but Ophiuchus and Lupus have very similar slopes. All the slopes, however, agree within 1$\sigma$. The slopes of the low-mass sample of Ophiuchus, Lupus and Cha-I are steeper than the single power-law slope and the high-mass sample slope, and the low-mass sample of Lupus does present a steeper slope than that of Ophiuchus. Therefore, we observe a steepening  of the slopes of the low-mass sample as observed in the $L_*-L\mathrm{_{acc}}$ relationship. As was the case for the accretion luminosity relation, Upper Scorpius does not follow the single power law or low-mass power-law trends found at younger ages. The high-mass sample slopes are consistent with each other and with linearity. In this case, if we do not consider the Lupus possible non-members, the single power-law slope of Lupus becomes shallower than in Ophiuchus.

The slopes of the low-mass sample found in Lupus and Cha-I (2.24 and 2.61, respectively) are consistently lower than those found in \citet{alcala17} and \citet{manara17cha} for the low-mass sample, respectively (4.58 and 6.45, respectively). The difference between the slopes can be explained by the different scale used to convert SpT to $T\mathrm{_{eff}}$.

We test which model is preferred using the same metrics as for the $L_*-L\mathrm{_{acc}}$ relationship (Sect.~\ref{results_lstar_lacc}). We find that the single power-law model is preferred in the Ophiuchus, Lupus, and Upper Scorpius regions. In Cha-I, we observe that the metrics show a slight preference of the double power-law model. The same conclusions are reached if the upper limits are not considered to calculate the metrics of the different fits.

Up to this point, we have found that the slopes of the accretion relationships ($L_*-L\mathrm{_{acc}}$ and $M_*-\dot{M}\mathrm{_{acc}}$) present a steepening in the 1-3 Myr age range (noting the caveat that the Lupus membership may affect this result in the $M_*-\dot{M}\mathrm{_{acc}}$ relationship). This steepening may be explained by a faster evolution into lower accretion rates at lower masses. We also found that the statistical significance of a double power law description of both relationships is similar to that of a single power law description. Given the simpler nature of the single power-law model, this may be the best description of both relationships. Therefore, we do not further discuss the possibility of these relationships being modeled by a double power law here. At the same time, the steepening of the single power-law models with age still implies that the evolution of the accretion rates is not the same across all masses.

\subsection{Relationship between stellar mass and protoplanetary disk mass}
\label{results_mstar_mdust}

\begin{figure*}[hbt!]
    \centering
    \includegraphics[width=\textwidth]
    {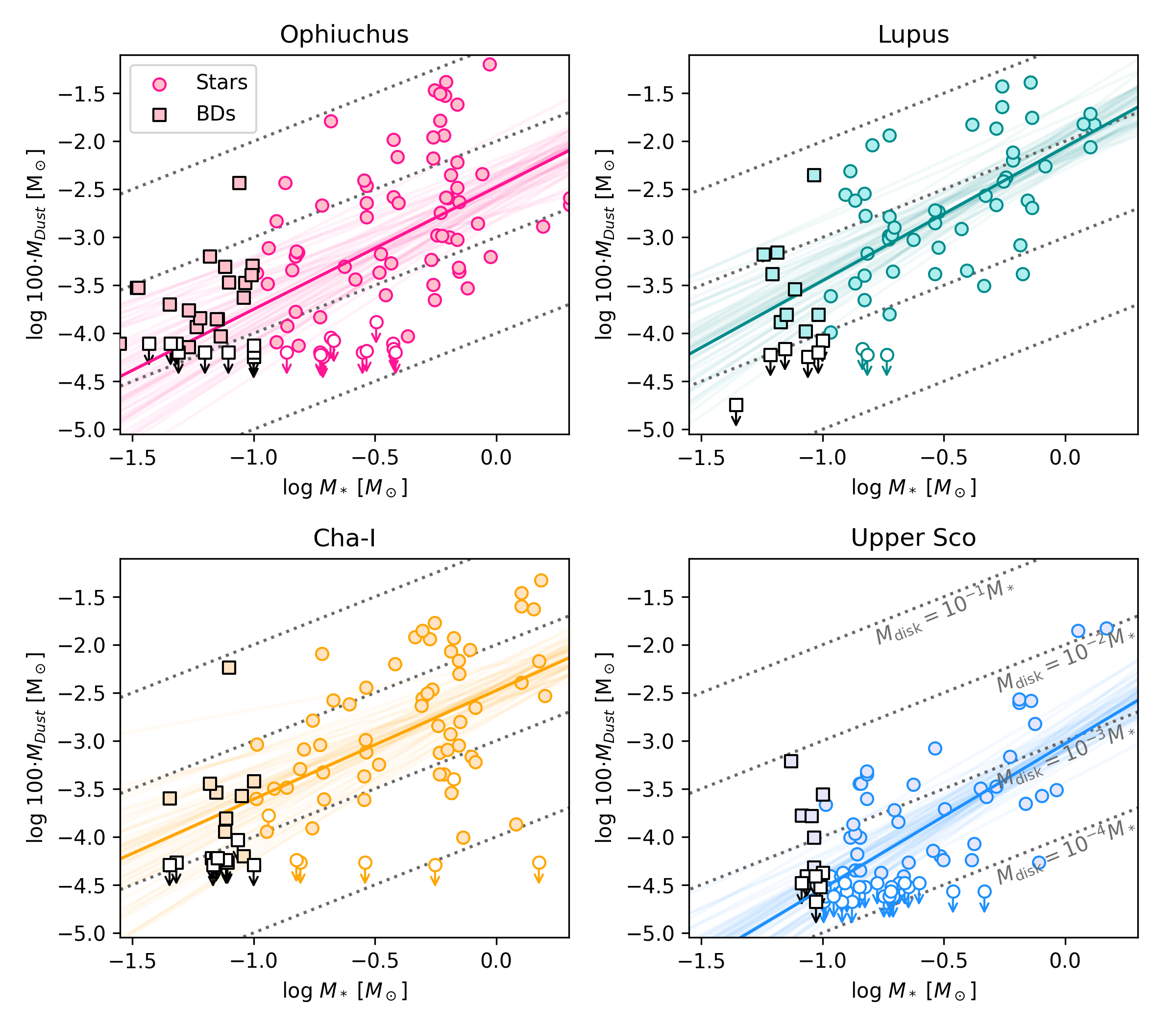}
    \caption{log$M_*-$log$M\mathrm{_{disk}}$ relationship for the four regions studied in this work: Ophiuchus (top left panel), Lupus (top right panel), Cha-I (bottom left panel), and Upper Scorpius (bottom right panel). The symbols follow the same color-coding as in Fig.~\ref{fig:lstar_lacc}. The stars are represented with circles and the BDs are represented as black squares. The solid lines represent the single power-law fits performed to the relationship in each region. The gray dotted lines mark the regions of constant $M\mathrm{_{disk}}/M_*$ ratio.}
    \label{fig:mstar_mdust}
\end{figure*}

In Fig.~\ref{fig:mstar_mdust}, we show the relationship between the stellar mass and the protoplanetary disk mass ($M\mathrm{_{disk}}$) of Ophiuchus, Lupus, Cha-I, and Upper Scorpius. We assume the typical 100 gas-to-dust ratio to convert the dust masses into total disk masses. In Fig.~\ref{fig:mstar_mdust}, we highlight the BDs (defined as sources with $M_*\leq$0.1 $M_\odot$) using black squares in order to evaluate whether the disks around BDs evolve similarly to those around stars (represented with circles). We observe that BDs in Ophiuchus, Lupus, and Cha-I mostly have $M\mathrm{_{disk}} < 10^{-2} M_*$, whereas all the regions (except Upper Scorpius) present a population of stars with $M\mathrm{_{disk}} > 10^{-2} M_*$ (see also Fig. 9). This seems to indicate that BDs rarely host very massive protoplanetary disks with $M\mathrm{_{disk}} > 10^{-2} M_*$ and that the relationship between both parameters is steeper than linear, similarly to previous studies \citep{pascucci16,testi22}. At the same time, the detection limit of the different ALMA surveys in these regions is located at $M\mathrm{_{disk}} \approx 10^{-4.7} M_\odot$ in Upper Scorpius and at $M\mathrm{_{disk}} \approx 10^{-4.3} M_\odot$ in the other regions. This detection limit represents the approximate value where most $M\mathrm{_{dust}}$ upper limit measurements lie. This detection limit is the reason why there are few BDs below the $M\mathrm{_{disk}} = 10^{-3} M_*$ line. In Upper Scorpius, we observe that most of the population has $M\mathrm{_{disk}} < 10^{-2} M_*$, although the population of BDs in Upper Scorpius remains largely unexplored.

\begin{figure*}[hbt!]
    \centering
    \includegraphics[width=\textwidth]{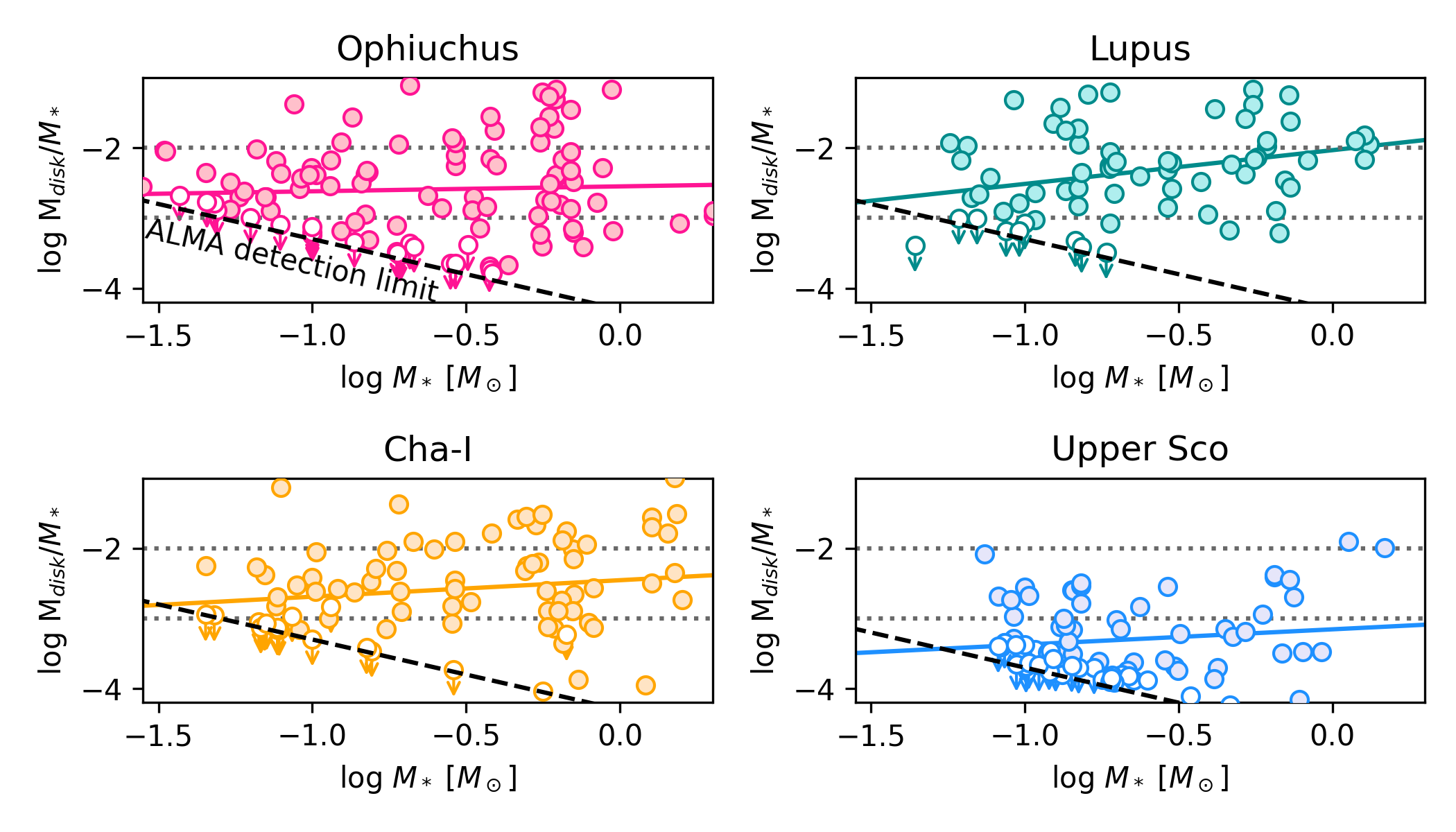}
    \caption{Ratio between disk mass and stellar mass as a function of the stellar mass for the Ophiuchus (top panel), Lupus (top middle panel), Cha-I (bottom middle panel), and Upper Scorpius (bottom panel) star-forming regions. The black dashed line represents the ALMA detection limit of the surveys in each region. The colored solid lines represent the power-law fit to the $M_*-M\mathrm{_{disk}}$ relationship performed in Sect.~\ref{results_mstar_mdust} and the gray dotted lines represent lines of constant $M\mathrm{_{disk}}/M_*$ ratio. The symbols follow the same color-coding as in Fig.~\ref{fig:lstar_lacc}.}
    \label{fig:mdust_over_mstar}
\end{figure*}

We performed a single power-law fit to the relationship in each region (see solid lines in Fig.~\ref{fig:mstar_mdust}). We observed that the correlation between both parameters is slightly steeper than linear in all the regions (slopes$=1.1-1.5$, see Table~\ref{tab:tab_slopes}). These slopes are shallower than those found in previous works for the same regions \citep{pascucci16,ansdell17,testi22}. This is consistent with the difference in the methodology used to derive masses between the works (see Sect.~\ref{results_literature}). If we exclude the BDs from the fit, the slopes are similar to those obtained using the entire mass range. In Fig.~\ref{fig:mdust_over_mstar}, we show the ratio between disk mass and stellar mass for the four regions studied in this work. We also show the best-fit power law to the $M_*-M\mathrm{_{disk}}$ relationship with solid lines. The black dashed line represents the detection limit of the ALMA surveys in each region (i.e., the locus of upper limit $M\mathrm{_{dust}}$ measurements in each region). We observe that the ALMA detection limit may be hiding the low-mass stars and BDs with even lower $M\mathrm{_{disk}}/M_*$ ratios, which would show the $M_*-M\mathrm{_{disk}}$ relationship to be steeper than what we have found.

\begin{figure}[hbt!]
    \centering
    \includegraphics[width=\textwidth/21*10]{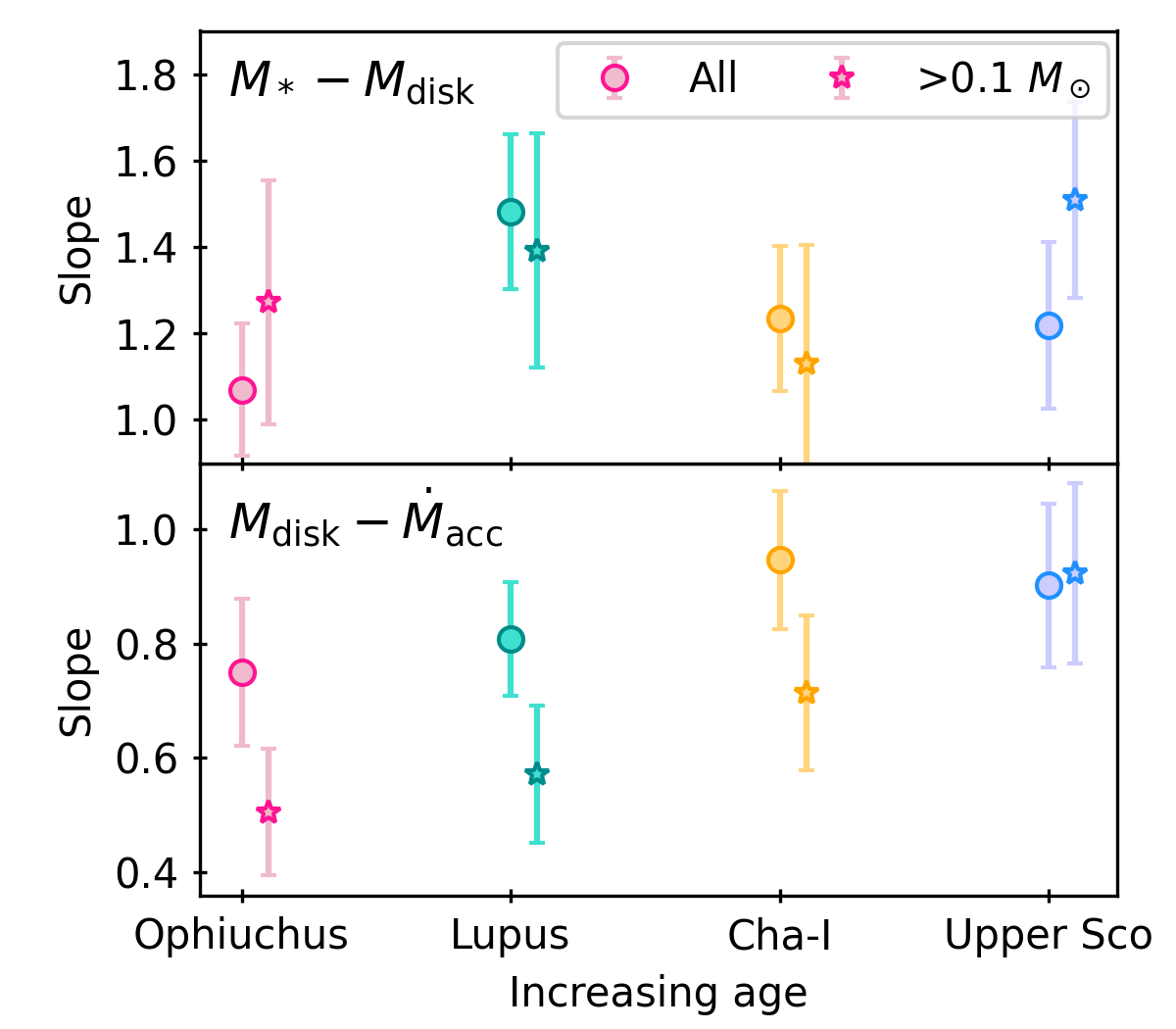}
    \caption{Slope of the power-law fits performed to the log$M_*-$log$M\mathrm{_{disk}}$ (top panel) and log$M\mathrm{_{disk}}-$log$\dot{M}\mathrm{_{acc}}$ (bottom panel) relationships for the four regions studied in this work. The circles represent the slopes of the power-law fit performed to the entire data set, and the stars that performed to the stellar population. A small offset in age in the slopes of each region was performed for visualization purposes. The symbols follow the same color-coding as in Fig.~\ref{fig:lstar_lacc}.}
    \label{fig:slope_age_disk}
\end{figure}

In the top panel of Fig.~\ref{fig:slope_age_disk} we show the slope of the $M_*-M\mathrm{_{disk}}$ relationship for each region. The circles represent the power-law fit performed to the entire mass range, and the star symbols represent the fit performed to the stellar population. We observe that independently of including the sub-stellar population or not, all the slopes are similar within 1$\sigma$. When the entire mass range is considered, Lupus presents the steepest slope, and when the sub-stellar population is not taken into account then it is Upper Scorpius that presents steeper slopes. Therefore, we do not observe any steepening of the $M_*-M\mathrm{_{disk}}$ relationship with age, which contrasts with the findings of previous works \citep{pascucci16,ansdell17}. The same results are found if we do not consider the possible non-members of Lupus.

\subsection{Relationship between protoplanetary disk mass and mass accretion rate}
\label{results_mdust_macc}

\begin{figure*}[hbt!]
    \centering
    \includegraphics[width=\textwidth]
    {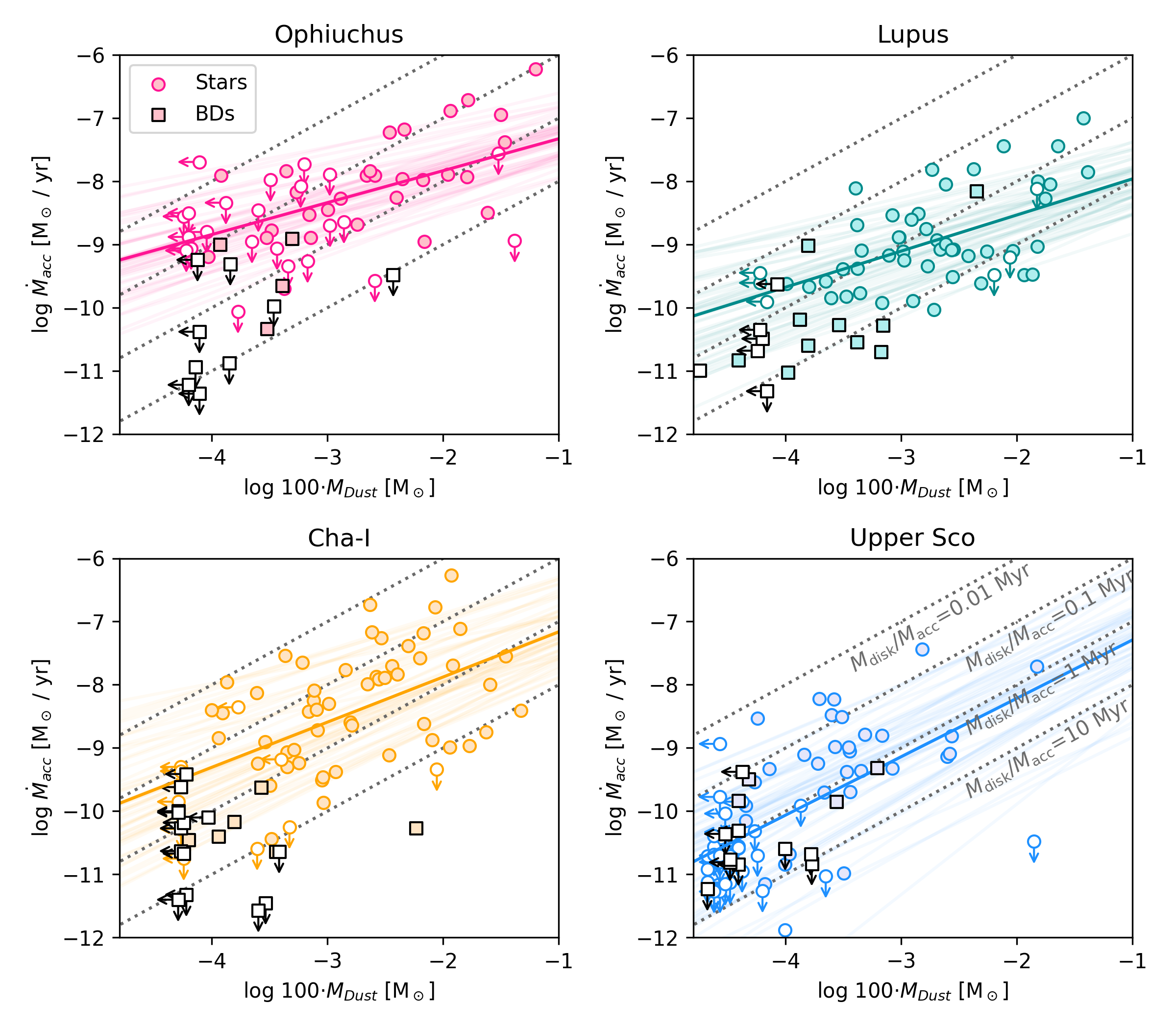}
    \caption{$M\mathrm{_{disk}}-\dot{M}\mathrm{_{acc}}$ relationship for the four regions studied in this work: Ophiuchus (top left panel), Lupus (top right panel), Cha-I (bottom left panel), and Upper Scorpius (bottom right panel). The symbols follow the same color-coding as in Fig.~\ref{fig:lstar_lacc}. The stars are represented with circles in colors and the BDs are represented as black squares. The solid lines represent the single power-law fits performed to the stellar population of each region. The gray dotted lines mark the regions of constant $\dot{M}\mathrm{_{acc}}/M\mathrm{_{disk}}$ ratio.}
    \label{fig:mdust_macc}
\end{figure*}

In Fig.~\ref{fig:mdust_macc}, we show the relationship between total disk mass and mass accretion rate for the Ophiuchus, Lupus, Cha-I, and Upper Scorpius star-forming regions. The dotted lines represent different constant $M\mathrm{_{disk}}/\dot{M}\mathrm{_{acc}}$ ratios (also known as accretion depletion timescale, \citealt{jones12,rosotti17}). Again we show the stars using circles and the BDs using black squares. We observe that the mass accretion rate increases with the protoplanetary disk mass and this correlation is consistent with being linear or sub-linear (i.e., slope$\leq 1$), in line with \citet{manara16disk} and \citet{mulders17}. We also observe that the BDs in Ophiuchus, Lupus, and Cha-I seem to present longer accretion depletion timescales \citep[the same behavior was previously reported in Lupus,][]{sanchis20}. However, the samples in all the regions contain a significant population of BDs with upper limits measurement of their dust masses, thus possibly consistent with shorter accretion depletion timescales. We performed a power-law fit to the stellar population (see solid lines in top panels of Fig.~\ref{fig:mdust_macc} and Table~\ref{tab:tab_slopes}) and observed that most of the BDs lie well below this line. The slopes we have derived range between 0.5 and 0.7 for Ophiuchus, Lupus, and Cha-I; whereas Upper Scorpius presents an almost linear slope. These slopes are also in agreement with the slopes found in \citet{testi22} using only $M_*>$0.15 $M_\odot$ sources, except for Ophiuchus where we find a shallower slope. We observe that Lupus is the only region with mean accretion depletion timescales longer than 1 Myr, although all the regions agree within 1 $\sigma$.

In the bottom panel of Fig.~\ref{fig:slope_age_disk} we show the slopes of the power-law fit performed to the stellar (star symbols) and entire population (circles) of the $M\mathrm{_{disk}}-\dot{M}\mathrm{_{acc}}$ relationship. We observe a steepening of the relationship in the entire age range when considering only the stellar population. The relationship in these regions becomes steeper by $\sim$0.2 units when the BDs are included in the fit, but the general trend is maintained (except for Upper Scorpius, but there are almost no substellar objects characterized in this region). If we do not consider the Lupus possible non-members in the fit performed to the entire population, the slope of the relationship in Lupus becomes very similar to that of Ophiuchus. At the same time, the power law performed to the stellar population maintains the same steepening trend when we are not considering the Lupus possible non-members.

\begin{figure}[hbt!]
    \centering
    \includegraphics[width=\textwidth/21*10]{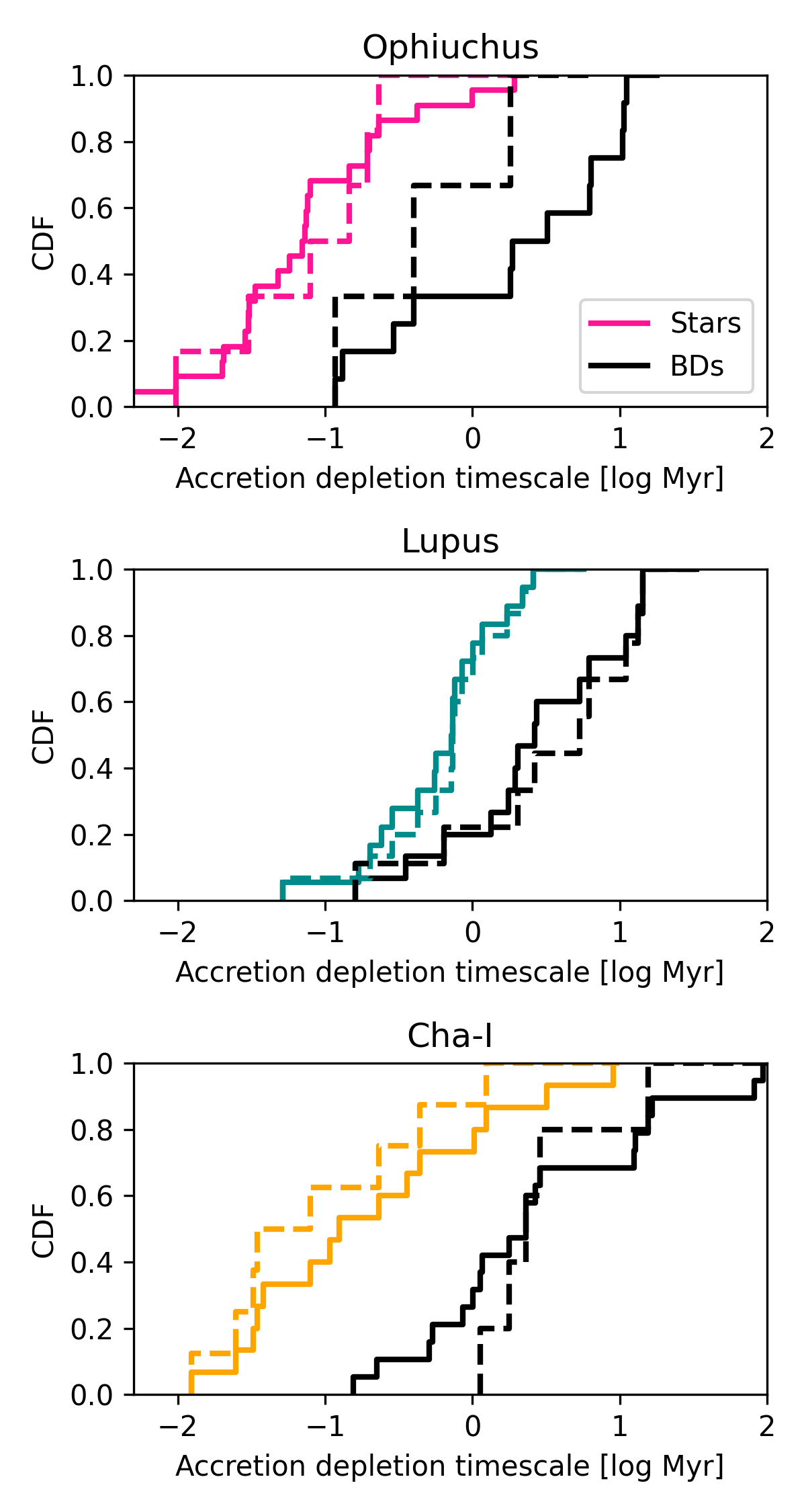}
    \caption{CDF of the accretion depletion timescale ($\dot{M}\mathrm{_{acc}}/M\mathrm{_{disk}}$) of Ophiuchus (top panel), Lupus (middle), and Cha-I (bottom). The black lines represent the CDF of BDs and the colored lines that of stars with $M\mathrm{_{disk}}<10^{-3}$ $M_\odot$ (color-coded by region as in Fig.~\ref{fig:lstar_lacc}). The dashed lines represent the same CDF excluding the upper limit measurements.}
    \label{fig:cdf_mdust_macc}
\end{figure}

We compared the distribution of the accretion depletion timescale of the BD and stellar populations with disk masses within the range of the BD disk masses (all disks with $M\mathrm{_{disk}}<10^{-3}$ $M_\odot$) in Ophiuchus, Lupus, and Cha-I. In Fig.~\ref{fig:cdf_mdust_macc} we show the cumulative distribution function (CDF) of the accretion depletion timescale for stars and BDs for the three regions. We used the Anderson-Darling test to evaluate whether both distributions are truly different by performing for each region a Monte Carlo simulation of the stellar and substellar distributions. We confirm that both distributions are indeed different in the three regions, with a probability $<$0.5\% of both distributions being sampled from the same underlying distribution. We conducted the same analysis excluding the upper limits (dashed lines) and we find that while in Lupus and Cha-I the result is very similar as before; in the case of Ophiuchus, the probability increases to 5.5\%. It is important to note that there is a very low number of BDs with detections in both accretion rate and protoplanetary dust mass in Ophiuchus. The same result is found if we do not consider the possible non-members in Lupus.

We found that the $M_*-M\mathrm{_{disk}}$ and $M\mathrm{_{disk}}-\dot{M}\mathrm{_{acc}}$ relationships present a different behavior than the $L_*-L\mathrm{_{acc}}$, $M_*-\dot{M}\mathrm{_{acc}}$ relationships. We found that the slopes of the $M_*-M\mathrm{_{disk}}$ relationship are similar in all the regions regardless of including BDs or not. On the other hand, the $M\mathrm{_{disk}}-\dot{M}\mathrm{_{acc}}$ relationship presents a steepening with age in the entire age range considered when we only evaluate the behavior of the stellar population, when including the BDs in the analysis the steepening stops at Cha-I, similarly to what we have found in the $L_*-L\mathrm{_{acc}}$ and $M_*-\dot{M}\mathrm{_{acc}}$ relationships. We have also found that BDs present longer accretion depletion timescales than stars in Ophiuchus, Lupus, and Cha-I.

\section{Discussion}
\label{xshoo_discussion}

\subsection{Accretion evolution of BDs}

Here, we discuss the accretion properties of BDs. In Sect.~\ref{results_mdust_macc} we found BDs to present larger $M\mathrm{_{disk}}/\dot{M}\mathrm{_{acc}}$ ratios (i.e., longer accretion depletion timescales) than average stars do. A similar result was first reported in \citet{sanchis20} for the Lupus star-forming region. \citet{sanchis20} argued that within the viscous disk evolution paradigm, a lower value of the $\alpha$ viscosity parameter would provide the lower accretion rates found for BDs. A lower viscosity would at the same time be explained by a global lower ionization rates in BDs \citep[e.g.,][]{mohanty05}.

Longer accretion depletion timescales could also be explained by a larger fraction of structured disks around BDs. Sources with transition disks have longer accretion depletion timescales than sources of the same disk mass \citep{najita07,najita15,manara16disk,mulders17,gangi22} similarly to what we have found for BDs \citep[see][for a theoretical explanation of this behavior in transition disks]{rosotti17}. In this scenario, our results would be naturally reproduced by a larger fraction of substructures in the disks of BDs. \citet{vandermarel21} found a stellar mass dependence of the fraction of structured disks, where high-mass stars are more prone to have structures in their disks. If this result holds for very low-mass sources, there would be a lower fraction of BDs with structured disks than low-mass stars, contrary to what would be needed to reproduce our result. A possible caveat is that the spatial resolution of the ALMA observations has not been able to uncover structures in the smallest disks. Currently there is a very small sample of low-mass stars and BDs where structured disks have been found \citep[see review by][]{pinilla22}. And most interestingly, if these structures found around low-mass stars and BDs are of planetary origin, they pose a strong challenge for planet formation models \citep{pinilla17,pinilla22}. Further high-angular-resolution observations of disks around BDs are needed to conclude how frequent are structures in disks around low-mass stars and BDs and whether they can explain this result \citep{pinilla21,kurtovic21,long23}.

\begin{figure}[hbt!]
    \centering
    \includegraphics[width=\textwidth/21*10]{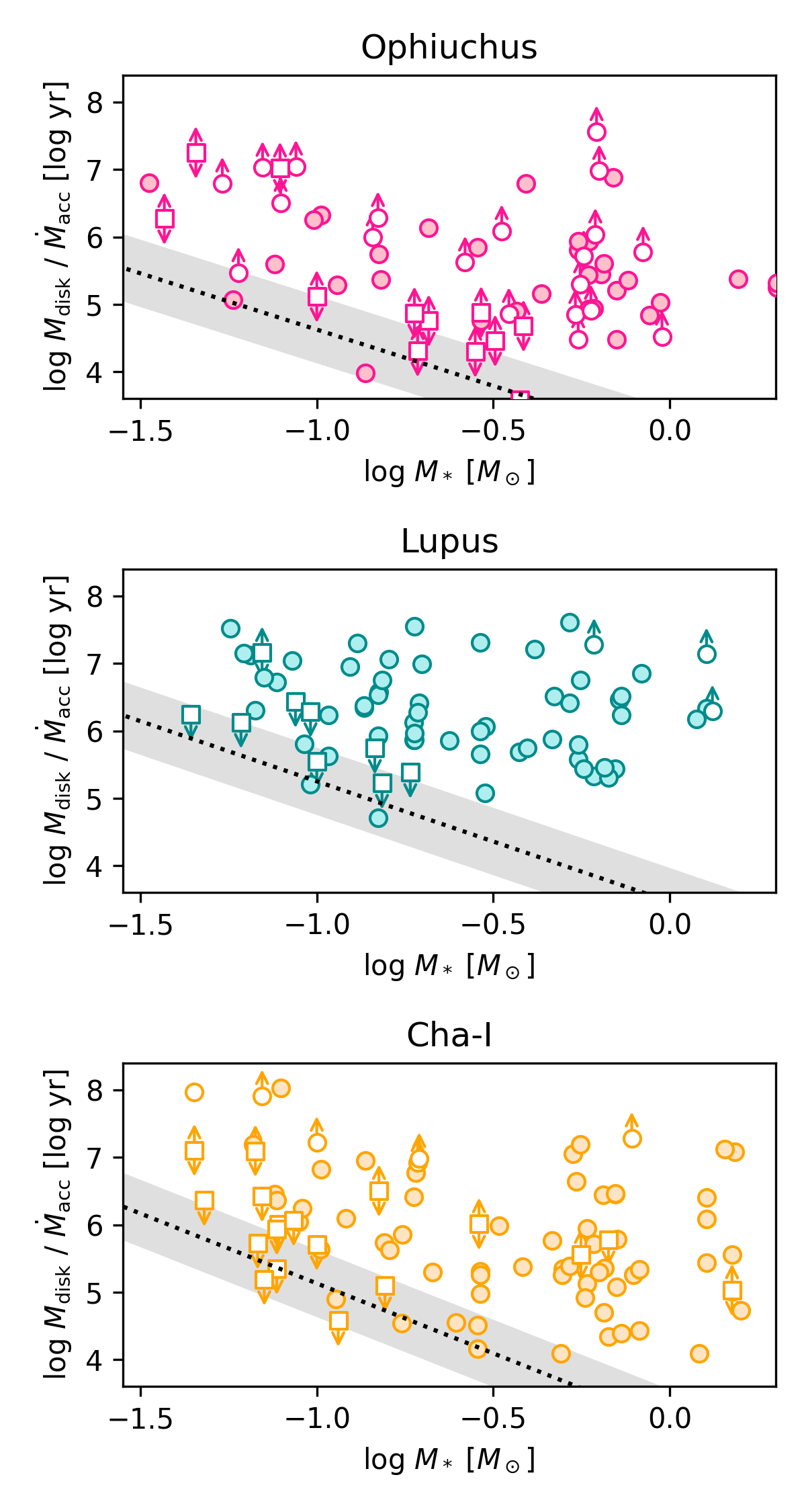}
    \caption{Accretion depletion timescale as a function of the stellar mass for the Ophiuchus (top panel), Lupus (middle panel), and Cha-I (bottom panel) star-forming regions. Empty squares represent sources without $M\mathrm{_{disk}}$ detections. The black dotted line and gray shaded area represent the estimated lower envelope and spread of depletion timescales based on the ALMA detection limit of these surveys. The symbols follow the same color-coding as in Fig.~\ref{fig:lstar_lacc}.}
    \label{fig:acc_timescale}
\end{figure}

An important caveat that needs to be addressed is that the longer depletion timescales of BDs could also be due to the detection limit of the ALMA observations. In order to test this, we estimated the lower envelope of accretion depletion timescales caused by the ALMA detection limit as a function of mass in each region. This estimate consists of a combination of the single power-law fits performed to the $M_*-\dot{M}\mathrm{_{acc}}$ relationships with an extra +0.5 dex (representing the upper locus of accretion rates in each region), and the ALMA detection limit, which is located at $M\mathrm{_{disk}} \approx 10^{-4.3} M_\odot$ in Ophiuchus, Lupus, and Cha-I. In Fig.~\ref{fig:acc_timescale} we show the accretion depletion timescale as a function of the stellar mass for Ophiuchus, Lupus, and Cha-I. We show the sources with ALMA non-detections with empty squares and a down-looking arrow. Accretion rate upper limit measurements are represented with empty circles and an up-looking arrow. The estimated lower envelopes of the accretion depletion timescales of each region are shown with a black dotted line. The gray shaded region represents a $\pm$0.5 dex spread. We observe that in Cha-I, and potentially in Lupus, the accretion depletion timescales of BDs may indeed be affected by the ALMA detection limits at masses $<$0.2 $M_\odot$. Therefore, deeper ALMA observations of disks around BDs are needed to unambiguously confirm this finding.

The accretion rate of BDs may also tell us about the mechanism by which these objects formed. \citet{stamatellos15} performed simulations of BDs formed by disk fragmentation in the protoplanetary disk of a higher mass star. They predicted that BDs formed this way would present accretion rates and protoplanetary disk masses independent of the central object's mass. Accretion rate measurements in substellar companions to T-Tauri stars have found high accretion rates \citep{zhou14,santamaria18,betti22}, in agreement with the predictions by \citet{stamatellos15}. On the other hand, the objects studied here are not companions and \citet{stamatellos15} did not model the effect that an ejection from the protoplanetary disk host would have on its accretion rates or disk masses. Therefore, the comparison with these predictions in isolated objects needs to be performed with caution. In the $L_*-L\mathrm{_{acc}}$ and $M_*-\dot{M}\mathrm{_{acc}}$ relationships, the new BDs do not stand out as outliers from the trend established by the low-mass stars (see Figs.~\ref{fig:lstar_lacc} and \ref{fig:mstar_macc}). However, the position of BDs in the $M_*-\dot{M}\mathrm{_{acc}}$ relationship of BDs is very sensitive to the $T\mathrm{_{eff}}-$SpT scale used. The two least massive BDs in Lupus (2MASS J15551027-3455045, J16085953-3856275) are located above the low-mass segmented power-law fit \citep[see also][]{majidi23}. In Cha-I, the low-mass BDs whose accretion measurement was presented here, may also deviate from the low-mass power-law fit. In the case of Cha-I, when using the \citet{herczeg14} $T\mathrm{_{eff}}-$SpT relationship for the low-mass BDs as well, these sources lie in much better agreement with the low-mass stars.

\citet{betti23} recently compiled a large number of mass accretion rate measurements of stars, BDs and planetary-mass companions. Using this dataset they found that BDs may follow a steeper correlation between $M_*$ and $\dot{M}\mathrm{_{acc}}$ than stars. They also found that the slope of this relationship for BDs undergoes a steepening with time, similarly to what we have found for the entire mass regime. The slopes of the $M_*-\dot{M}\mathrm{_{acc}}$ relationship found in \citet{betti23} agree well with the values we have found in this work for the same age ranges. However, we do not find the BDs to follow a distinct steeper correlation between the $M_*$ and $\dot{M}\mathrm{_{acc}}$, not even if we consider the \citet{herczeg14} $T\mathrm{_{eff}}-$SpT relationship for the low-mass BDs as well (as discussed above). It is important to note that \citet{betti23} found the slope of the BD regime to undergo the most important steepening at ages older than 3 Myr, which is above the oldest age of the regions we have studied into the BD regime in this work. Another interesting result from \citet{betti23}, is that they found that the accretion rates of BDs derived from the H$\alpha$ line luminosity may be overestimated compared to the measurement using the UV continuum excess. We cannot test this result on the X-Shooter sample presented in this work, because no signal was detected in the UVB arm in any of the BDs. However, if this overestimation of the accretion rates is real, it would imply that the slopes of the $M_*-\dot{M}\mathrm{_{acc}}$ and $M\mathrm{_{disk}}-\dot{M}\mathrm{_{acc}}$ relationships in the BD regime may actually be steeper than what we have found. It would also imply that BDs would have longer accretion depletion timescales than what we have found, making them even more different than stars.

\subsection{Implications for the understanding of disk evolution}

There is general agreement that protoplanetary disks around high-mass stars ($\geq$1.5 $M_\odot$) evolve faster than their low-mass counterparts, based on the study of the protoplanetary disk fraction (i.e., sources with excess emission caused by warm dust in the inner part of the disk over total number of members of a young cluster) in young clusters \citep{carpenter06,ribas15}. The low-mass regime ($<$1 $M_\odot$) presents no clear trend with mass in the fraction of disk bearing stars at young ages \citep[$\sim$1-4 Myr;][]{luhman10,monin10}. However, the older Upper Scorpius presents hints of larger disk fractions in low-mass stars and BDs \citep{scholz07,luhman12usco,luhman22disk,dawson13}, as well as in the even older ($\sim$20 Myr) associations in the Sco-Cen complex \citep{luhman22disk}. A similar mass-dependent disk fraction was also found in the $\sim$5 Myr old $\lambda$ Ori star-forming region \citep{bayo12}. These results may indicate that very low-mass sources have longer-lived protoplanetary disks. At the same time, we observed that the correlation between the stellar mass and the protoplanetary disk mass is slightly larger than one (in line with previous results, e.g., \citealt{pascucci16}). Consequently, the ratio between disk mass and stellar mass decreases with the mass of the central object (see Fig.~\ref{fig:mdust_over_mstar}). And in this work, we have found that very low-mass stars and BDs have a faster evolution in terms of their accretion rates, leading to lower values than their higher mass counterparts. Such a result is in agreement with the results of \citet{betti23}, although we have found that this phenomena is already at work at even earlier ages ($\leq$3 Myr).

Overall, these results pose an apparently contradictory scenario: very low-mass sources evolve faster into lower accretion rates, they may possess lower disk masses but they also may have longer-lived disks. This discrepancy may be alleviated if the evolutionary timescale of the dusty inner disk with respect to the timescale of the accretion processes is different. Transition disks present a very different behavior compared to low-mass sources: transition disks have no dust in the inner part of the disk, but there is ongoing accretion \citep[see review by][and references therein]{vandermarel23}. There is a number of sources with evidence of dust in their inner disk, but no appearance of accretion \citep[see][and references therein]{manara17cha}, such as the sources presented in this work, which have had their infrared excess detected and whose accretion diagnostics are compatible with chromospheric activity.

In the context of viscously evolving disks, the strong dependence of the mass accretion rates with stellar masses ($\dot{M}\mathrm{_{acc}} \propto M_*^2$) has been reproduced in different works \citep{alexander06,hartmann06,dullemond06,somigliana22}. \citet{alexander06} studied the $M_*-\dot{M}\mathrm{_{acc}}$ relationship in the viscous evolution paradigm assuming that this relationship is set by a variation in the initial conditions with the stellar mass, in particular, so that the ``initial'' disk size increases with decreasing stellar mass. Such a scenario resulted in BDs presenting a slower evolution in terms of the accretion rates (i.e., a longer viscous timescale), which is contrary to the results of our study. On the other hand, \citet{hartmann06} reproduced the shape of the $M_*-\dot{M}\mathrm{_{acc}}$ relationship in the context of viscous evolution, invoking the notion that the disk masses decrease steeply with the stellar mass. In these models, the authors found that viscous evolution is faster in lower mass stars (i.e., a shorter viscous timescale), similarly to what we have found. \citet{somigliana22} found that the slope of the $M_*-\dot{M}\mathrm{_{acc}}$ relationship would undergo an evolutionary steepening when the $M_*-M\mathrm{_{disk}}$ relationship is initially steeper than the $M_*-\dot{M}\mathrm{_{acc}}$ relationship. The slope of the relationship between the disk mass and the accretion rate is also naturally reproduced by viscous models \citep{dullemond06,mulders17,lodato17}. At the same time, viscous models predict that the spread of this relationship becomes smaller with age \citep{lodato17}. \citet{manara20} found that the $M_*-M\mathrm{_{disk}}$ relationship of Upper Scorpius presented a similar spread as younger regions, contrary to predictions of viscous evolution. \citet{sellek20} studied this issue including a modeling of the evolution of dust and found that this discrepancy can be reconciled when taking into account that disk masses are measured through the continuum flux. Another possibility is that the binary population may have biased the observed accretion depletion timescales toward lower values \citep{zagaria22}. In any case, a model that explains not only the steepening of the accretion rate scaling relationships but also that of the $M\mathrm{_{disk}}-\dot{M}\mathrm{_{acc}}$ relationship and the corresponding spreads has not been put forward so far.

Current MHD wind disk models have no particular prediction of the shape and evolution of these relationships. However, the initial conditions can be tailored to reproduce the accretion and disk mass relationships \citep[e.g.,][]{mulders17,tabone22}. These models also naturally reproduce the spread in the $M\mathrm{_{disk}}-\dot{M}\mathrm{_{acc}}$ relationship. Recently, \citet{somigliana23} showed that MHD wind disk models predict a smaller decrease with time of the spread in the $M\mathrm{_{disk}}-\dot{M}\mathrm{_{acc}}$ relationship than purely viscous models.

\section{Summary and conclusions}
\label{xshoo_summary}

In this work, we present the analysis of the protoplanetary disk and accretion scaling relations in four different star-forming regions: Ophiuchus ($\sim$1 Myr), Lupus ($\sim$2 Myr), Cha-I ($\sim$3 Myr), and Upper Scorpius (5-12 Myr). This work included 26 new X-Shooter spectra of low-mass stars and BDs in Ophiuchus, Cha-I and Upper Scorpius. We derived the physical and accretion properties of these objects, and combined the new measurements with those from the literature. The combined sample of sources was analyzed using a homogeneous methodology.

We analyzed the shape and time evolution of the disk scaling relationships ($L_*-L\mathrm{_{acc}}$, $M_*-\dot{M}\mathrm{_{acc}}$, $M_*-M\mathrm{_{disk}}$, $M\mathrm{_{disk}}-\dot{M}\mathrm{_{acc}}$) in the four different star-forming regions studied here. We found that the accretion relationships ($L_*-L\mathrm{_{acc}}$, $M_*-\dot{M}\mathrm{_{acc}}$) are reproduced by a single or a double power law with a similar statistical significance. Since the single power-law model is simpler, we find this model to be a better description of both relationships. We found that the slopes of the accretion relationships undergo a steepening with time in the 1-3 Myr age range (i.e., between Ophiuchus, Lupus, and Cha-I). For the $M\mathrm{_{disk}}-\dot{M}\mathrm{_{acc}}$ relationship we found that the steepening extends down to the age of Upper Scorpius if only the stellar population is considered (the BD population of Upper Scorpius has almost no measurements of their protoplanetary disk dust content). This result is most possibly driven by a faster evolution of the low-mass stars into lower accretion rates while keeping an equally lower $M\mathrm{_{disk}}$. The single power-law slope of the accretion relationships in Upper Scorpius does not follow the trend seen at younger ages, which may indicate that all sources have evolved into lower accretion rates. However, it is important to note that the accretion rate and disk mass measurements in Upper Scorpius are very incomplete and do not extend to the BD regime.

We found the $M_*-M\mathrm{_{disk}}$ relationship to be steeper than linear, but we did not find any evidence for an evolutionary behavior of the slopes of this relationship as previously proposed \citep{pascucci16,ansdell17}. However, the sensitivity limit of the ALMA surveys in these regions needs to be increased to better understand the time evolution of this relationship better. We also found that the BD population of the regions studied in this work present longer accretion depletion timescales than the stellar population, which was confirmed via an Anderson-Darling test between the BD and stellar populations. A possible caveat of this result is the current detection limit of protoplanetary disk dust observations.

Overall, this field would immensely benefit from an increase in the number of low-mass stars and BDs with their accretion rates measured, especially in the case of the Ophiuchus and Upper Scorpius star-forming regions. Subsequent deep ALMA observations of disks around BDs may be able to confirm whether BDs present longer accretion depletion timescales, while high angular resolution ALMA observations of BDs will also help to constrain our understanding of the impact of substructures in disks around BDs.

\begin{acknowledgements}
   V.A-A. acknowledges funding by the Science and Technology Foundation of Portugal (FCT), grants No. IF/00194/2015, PTDC/FIS-AST/28731/2017, UIDB/00099/2020 and SFRH/BD/143433/2019. V.A-A acknowledges ESO for their support during the initial six months of this work. Funded by the European Union (ERC, WANDA, 101039452). Views and opinions expressed are however those of the author(s) only and do not necessarily reflect those of the European Union or the European Research Council Executive Agency. Neither the European Union nor the granting authority can be held responsible for them. This work was partly supported by the Italian Ministero dell’Istruzione, Universit\`{a} e Ricerca through the grant Progetti Premiali 2012-iALMA (CUP C52I13000140001), by the Deutsche Forschungsgemeinschaft (DFG, German Research Foundation) - Ref no. 325594231 FOR 2634/2 TE 1024/2-1, by the DFG Cluster of Excellence Origins (www.origins-cluster.de). This project has received funding from the European Union’s Horizon 2020 research and innovation program under the Marie Skłodowska- Curie grant agreement No 823823 (DUSTBUSTERS), and from the European Research Council (ERC) via the ERC Synergy Grant ECOGAL (grant 855130). K.M. acknowledges support from the Fundação para a Ciência e a Tecnologia (FCT) through the CEEC-individual contract 2022.03809.CEECIND and research grants UIDB/04434/2020 and UIDP/04434/2020. JMA acknowledge financial support from the project PRIN-INAF 2019 "Spectroscopically Tracing the Disk Dispersal Evolution (STRADE) and the Large Grant INAF 2022 "YSOs Outflows, Disks and Accretion" (YODA).
\end{acknowledgements}

\bibliographystyle{aa} 
\bibliography{bds_acc}

\begin{appendix}

\section{Observing log and literature parameters}

In Table~\ref{tab:tab_obs}, we present the observing log of the X-Shooter observations presented in Sect.~\ref{xshoo_observations}. In Table~\ref{tab:tab_lit_params}, we show the coordinates, distance, region to which they belong and spectroscopic parameters from the literature of the observed sample.

\begin{table*}[hbt!]
    \caption{X-Shooter observing log of the sample presented in Sect.~\ref{xshoo_observations}.}
    \begin{center}
        \resizebox{\textwidth}{!}{
        \begin{tabular}{ l | c | c c c | c c c | c c c }
            \hline\hline
            Name & Date & \multicolumn{3}{c|}{Slit width ["$\times$ 11"]} & \multicolumn{3}{c|}{Exp. time [Nexp$\times$t(s)]} & \multicolumn{3}{c}{$S/N$ at $\lambda=$} \\ 
             & & UVB & VIS & NIR & UVB & VIS & NIR & 700 & 855 & 1300 \\ 
            \hline
            CHSM 12653 & 2021-04-30 & 1.0 & 0.9 & 0.9 & 4$\times$560 & 4$\times$650 & 4$\times$700 & 3 & 14 & 19 \\ 
            2MASS J11084952-7638443 & 2021-04-09 & 1.0 & 0.9 & 0.9 & 4$\times$660 & 4$\times$750 & 4$\times$750 & 1 & 4 & 13 \\ 
            2MASS J11112249-7745427 & 2021-04-30 & 1.0 & 0.9 & 0.9 & 4$\times$660 & 4$\times$750 & 4$\times$750 & 1 & 5 & 15 \\ 
            2MASS J11114533-7636505 & 2021-05-02 & 1.0 & 0.9 & 0.9 & 4$\times$610 & 4$\times$700 & 4$\times$750 & 2 & 8 & 17 \\ 
            2MASS J11122250-7714512 & 2021-04-29 & 1.0 & 0.9 & 0.9 & 4$\times$660 & 4$\times$750 & 4$\times$750 & 0 & 3 & 12 \\ 
            2MASS J16031329-2112569 & 2021-06-28 & 1.0 & 0.4 & 0.4 & 4$\times$520 & 4$\times$430 & 4$\times$500 & 31 & 88 & 80 \\ 
            2MASS J16052661-1957050 & 2021-06-28 & 1.0 & 0.4 & 0.4 & 4$\times$450 & 4$\times$360 & 4$\times$500 & 38 & 100 & 102 \\ 
            2MASS J16053215-1933159 & 2021-06-28 & 1.0 & 0.4 & 0.4 & 4$\times$550 & 4$\times$460 & 4$\times$600 & 16 & 53 & 44 \\ 
            2MASS J16060061-1957114 & 2021-06-29 & 1.0 & 0.4 & 0.4 & 4$\times$450 & 4$\times$360 & 4$\times$500 & 54 & 127 & 132 \\ 
            PGZ2001 J160702.1-201938 & 2021-06-29 & 1.0 & 0.4 & 0.4 & 4$\times$550 & 4$\times$460 & 4$\times$600 & 20 & 64 & 60 \\ 
            2MASS J16083455-2211559 & 2021-06-29 & 1.0 & 0.4 & 0.4 & 4$\times$550 & 4$\times$460 & 4$\times$600 & 26 & 76 & 64 \\ 
            2MASS J16101888-2502325 & 2021-05-04 & 1.0 & 0.4 & 0.4 & 4$\times$520 & 4$\times$430 & 4$\times$500 & 43 & 103 & 94 \\ 
            2MASS J16102819-1910444 & 2021-09-04 & 1.0 & 0.4 & 0.4 & 4$\times$650 & 4$\times$560 & 4$\times$700 & 20 & 66 & 50 \\ 
            2MASS J16145928-2459308 & 2021-05-04 & 1.0 & 0.4 & 0.4 & 4$\times$520 & 4$\times$430 & 4$\times$500 & 46 & 110 & 100 \\ 
            2MASS J16151239-2420091 & 2021-07-05 & 1.0 & 0.4 & 0.4 & 4$\times$650 & 4$\times$560 & 4$\times$700 & 21 & 66 & 42 \\ 
            2MASS J16181618-2619080 & 2021-05-04 & 1.0 & 0.4 & 0.4 & 4$\times$550 & 4$\times$460 & 4$\times$600 & 22 & 76 & 82 \\ 
            CRBR 2317.5-1729 & 2021-08-30 & 1.0 & 0.9 & 0.9 & 4$\times$660 & 4$\times$750 & 4$\times$750 & 0 & 0 & 9 \\ 
            CRBR 2322.3-1143 & 2021-08-31 & 1.0 & 0.9 & 0.9 & 4$\times$610 & 4$\times$700 & 4$\times$750 & 0 & 1 & 10 \\ 
            ISO-Oph042 & 2022-07-13 & 1.0 & 0.9 & 0.9 & 4$\times$630 & 4$\times$700 & 4$\times$750 & 1 & 8 & 45 \\ 
            GY92 80 & 2021-07-18 & 1.0 & 0.9 & 0.9 & 4$\times$610 & 4$\times$700 & 4$\times$750 & 1 & 7 & 29 \\ 
            GY92 90 & 2021-08-05 & 1.0 & 0.9 & 0.9 & 4$\times$660 & 4$\times$750 & 4$\times$750 & 0 & 1 & 9 \\ 
            SONYC RhoOph-6 & 2021-09-03 & 1.0 & 0.9 & 0.9 & 4$\times$770 & 4$\times$860 & 4$\times$900 & 0 & 0 & 2 \\ 
            GY92 264 & 2021-09-01 & 1.0 & 0.9 & 0.9 & 4$\times$660 & 4$\times$750 & 4$\times$750 & 16 & 61 & 113 \\ 
            CFHTWIR-Oph 77 & 2022-06-11 & 1.0 & 0.9 & 0.9 & 4$\times$930 & 4$\times$1020 & 4$\times$1050 & 0 & 0 & 1 \\ 
            BKLT J162736-245134 & 2021-07-17 & 1.0 & 0.9 & 0.9 & 4$\times$660 & 4$\times$750 & 4$\times$750 & 0 & 0 & 1 \\ 
            BKLT J162848-242631 & 2021-07-07 & 1.0 & 0.9 & 0.9 & 4$\times$560 & 4$\times$650 & 4$\times$700 & 4 & 22 & 26 \\ 
            \hline 
        \end{tabular}
        }
    \end{center}
    \label{tab:tab_obs}
\end{table*}

\begin{table*}
    \caption{Coordinates and literature parameters of the observed sample.}
    \begin{center}
        \begin{tabular}{l c c c c c c}
            \hline\hline
            Object/other name & RA(2000) & Dec(2000) & Region & SpT & Av & References \\ 
            \hline
            CHSM 12653 & 11:08:29.270 & -77:39:19.836 & Cha-I & M7.25 & 0.0 & 1 \\ 
            2MASS J11084952-7638443 & 11:08:49.524 & -76:38:44.350 & Cha-I & M8.75 & 0.0 & 1 \\ 
            2MASS J11112249-7745427 & 11:11:22.499 & -77:45:42.710 & Cha-I & M8.25 & 0.0 & 1 \\ 
            2MASS J11114533-7636505 & 11:11:45.336 & -76:36:50.533 & Cha-I & M8 & 1.2 & 1 \\ 
            2MASS J11122250-7714512 & 11:12:22.509 & -77:14:51.237 & Cha-I & M9.25 & 0.0 & 1 \\ 
            2MASS J16031329-2112569 & 16:03:13.297 & -21:12:56.905 & Upper Sco & M4.75 & 0.8 & 2 \\ 
            2MASS J16052661-1957050 & 16:05:26.611 & -19:57:05.050 & Upper Sco & M4.5 & 0.2 & 2 \\ 
            2MASS J16053215-1933159 & 16:05:32.151 & -19:33:15.994 & Upper Sco & M4.75 & 0.9 & 2 \\ 
            2MASS J16060061-1957114 & 16:06:00.615 & -19:57:11.455 & Upper Sco & M4 & 0.9 & 2 \\ 
            PGZ2001 J160702.1-201938 & 16:07:02.118 & -20:19:38.769 & Upper Sco & M5 & 0.8 & 2 \\ 
            2MASS J16083455-2211559 & 16:08:34.552 & -22:11:55.917 & Upper Sco & M4.5 & 0.3 & 2 \\ 
            2MASS J16101888-2502325 & 16:10:18.881 & -25:02:32.546 & Upper Sco & M4.75 & 0.3 & 2 \\ 
            2MASS J16102819-1910444 & 16:10:28.195 & -19:10:44.486 & Upper Sco & M4 & 0.5 & 2 \\ 
            2MASS J16145928-2459308 & 16:14:59.287 & -24:59:30.804 & Upper Sco & M4.25 & 1.0 & 2 \\ 
            2MASS J16151239-2420091 & 16:15:12.393 & -24:20:09.121 & Upper Sco & M4 & 0.3 & 2 \\ 
            2MASS J16181618-2619080 & 16:18:16.185 & -26:19:08.065 & Upper Sco & M4.5 & 2.1 & 2 \\ 
            CRBR 2317.5-1729 & 16:26:18.982 & -24:24:14.259 & Ophiuchus & M6 & 24.4 & 3 \\ 
            CRBR 2322.3-1143 & 16:26:23.814 & -24:18:29.001 & Ophiuchus & M5.5 & 8.5 & 3 \\ 
            ISO-Oph042 & 16:26:27.809 & -24:26:41.82 & Ophiuchus & M5 & 6.2 & 4 \\ 
            GY92 80 & 16:26:37.807 & -24:39:03.196 & Ophiuchus & M5.5 & 6.8 & 3 \\ 
            GY92 90 & 16:26:39.917 & -24:22:33.427 & Ophiuchus & M8.25 & 7.6 & 3 \\ 
            SONYC RhoOph-6 & 16:27:05.926 & -24:18:40.215 & Ophiuchus & M8 & 6.8 & 3 \\ 
            GY92 264 & 16:27:26.581 & -24:25:54.386 & Ophiuchus & M8 & 0.3 & 3 \\ 
            CFHTWIR-Oph 77 & 16:27:25.644 & -24:37:28.516 & Ophiuchus & M9.75 & 10 & 5 \\ 
            BKLT J162736-245134 & 16:27:36.611 & -24:51:36.09 & Ophiuchus & L0 & 2.1 & 3 \\ 
            BKLT J162848-242631 & 16:28:48.704 & -24:26:31.794 & Ophiuchus & M6.25 & 1.9 & 3 \\ 
            \hline 
        \end{tabular}
    \tablebib{(1) \citet{esplin17}; (2) \citet{luhman18}; (3) \citet{esplin20}; (4) \citet{wilking05}; (5) \citet{alvesoliveira12}.}
    \end{center}
    \label{tab:tab_lit_params}
\end{table*}

\section{Membership of the sources used in this work}
\label{lupus_membership}

In this section, we evaluate the membership of all the sources used in this work to their respective star-forming regions. We compared their \textit{Gaia} DR3 kinematics \citep{gaiadr3} and location to that of known members of the different regions \citep{luhman22}. To restrict the comparison solely to the sources with reliable \textit{Gaia} astrometry, we made a cut in RUWE$<$2.5 \citep{fitton22}. We found that the sources of all regions except for Lupus agree well with the kinematics and location of known members of each region. In the case of Lupus, we found a number of sources whose kinematics and location could also be explained with membership to another of the groups of the Scorpius-Centaurus complex. Therefore, we made a more detailed comparison of the location and kinematics of these sources with the catalog of Lupus members from \citet{luhman20_lupus}, and that of Upper Scorpius, Upper Centaurus Lupus (UCL), and Lower Centaurus Crux (LCC) from \citet{luhman22}. The main flags for possible non membership to Lupus that we find to be the most useful are:

\begin{itemize}
    \item Proper motion in right ascension $<$ -15 "/yr: Below this value, membership to UCL would be much more probable than membership to Lupus.
    \item Located at a distance $<$ 147 pc (or parallax $>$ 6.8"): at shorter distances, the parameter space is dominated by members of UCL and Upper Scorpius.
    \item Located off clouds: the location of the Lupus clouds completely overlaps with the extent of UCL members.
\end{itemize}

Based on the evaluation of these aspects we find the following sources with reliable \textit{Gaia} DR3 astrometry to be possible non-members of Lupus:

\begin{itemize}
    \item Sources 2MASS J15445789-3423392, 2MASS J15592523-4235066, 2MASS J15414081-3345188, 2MASS J16011870-3437332, 2MASS J15383733-3422022, 6010590577947703936, 2MASS J15414827-3501458, UCAC4273-083363 and 6014269268967059840 all have proper motions in right ascension $<$ -15 "/yr, which is in better agreement  with a membership to UCL. Only one of these sources is located in the cloud (2MASS J15592523-4235066).
    \item SSTc2d J160828.1-391310 is on cloud but its proper motion in right ascension is close -15 "/yr and its proper motion in declination is also just outside the locus of the known Lupus members.
    \item 2MASS J16134410-3736462 is slightly off cloud, it has proper motions in right ascension -13.5 "/yr and it is located at the distance of known Lupus members.
    \item 2MASS J15361110-3444473 is off cloud and has slightly discrepant proper motions that could make it a UCL member (larger proper motions in declination). It is also located slightly further away from the cluster, however, within the cluster's distance in 1-2$\sigma$.
    \item 2MASS J16081497-3857145 is on cloud but has a proper motion in declination that is much lower than other Lupus members, which could make it rather a member of Upper Scorpius.
\end{itemize}

\begin{figure*}[hbt!]
    \centering
    \includegraphics[width=\textwidth]{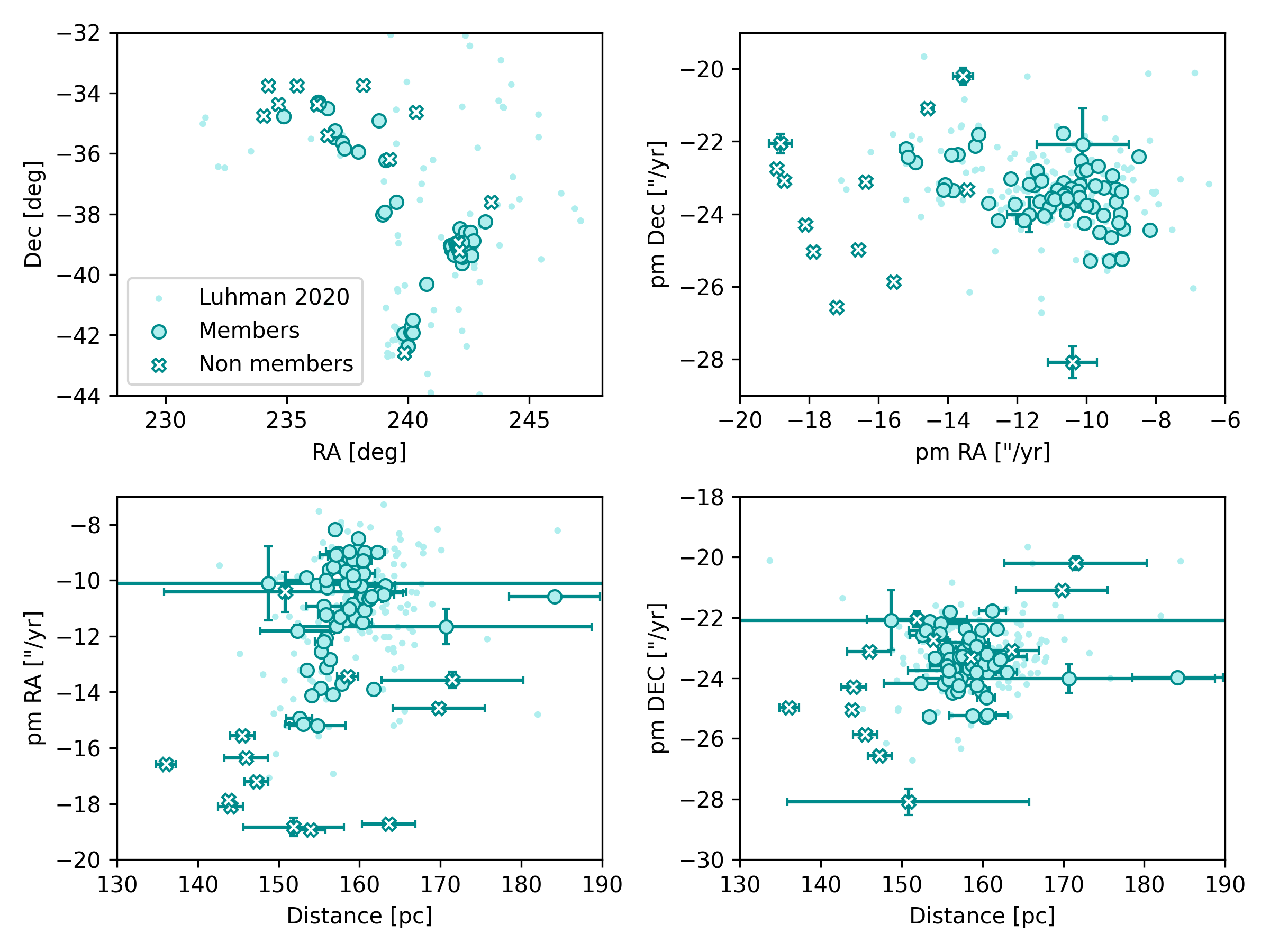}
    \caption{Location and kinematics of the Lupus star-forming region. Top-left panel shows the on-sky location, top-right panel shows the kinematics, bottom-left panel shows the comparison between the distance, and the proper motion in right ascension, and the bottom right panel shows the comparison between the distance and the proper motion in declination. Adopted members are shown with cyan circles and the possible non members are shown with white-filled crosses. Members of \citet{luhman20_lupus} are shown with small circles.}
    \label{fig:lupus_membership}
\end{figure*}

Figure~\ref{fig:lupus_membership} shows the location and kinematics of the adopted members and possible non-members of Lupus together with the Lupus members of \citet{luhman20_lupus}.

\section{Power law fit results}

In Table~\ref{tab:tab_slopes}, we present the results of the power-law fits performed to the four relationships studied in Sect.~\ref{xshoo_results}. For the $M_*-M\mathrm{_{disk}}$ and $M\mathrm{_{disk}}-\dot{M}\mathrm{_{acc}}$ relationships we also provide the results of the power-law fits performed to the stellar population ($>$0.1 $M_\odot$). All the fits were performed on a \textit{log} scale of both quantities involved in each fit.

\begin{table}
    \caption{Parameters of the power-law fits performed in this work. Note: all the power-law fits have been performed on the logarithm of both variables involved in each relationship.}
    \begin{center}
        \begin{tabular}{l c c c}
            \hline\hline
            Relationship & Region & Slope & Intercept \\ 
            \hline
            $L_*-L\mathrm{_{acc}}$ & Ophiuchus & 1.19$\pm$0.08 & -1.46$\pm$0.08 \\ 
             & Lupus & 1.52$\pm$0.12 & -1.46$\pm$0.14 \\ 
             & (only members) & 1.46$\pm$0.15 & -1.47$\pm$0.14 \\ 
             & Cha-I & 1.76$\pm$0.11 & -1.1$\pm$0.12 \\ 
             & Upper Sco & 1.5$\pm$0.17 & -1.65$\pm$0.2 \\ 
            \hline
            $M_*-\dot{M}\mathrm{_{acc}}$ & Ophiuchus & 1.67$\pm$0.18 & -7.76$\pm$0.14 \\ 
             & Lupus & 1.79$\pm$0.22 & -8.27$\pm$0.18 \\ 
             & (only members) & 1.41$\pm$0.22 & -8.35$\pm$0.16 \\ 
             & Cha-I & 2.08$\pm$0.18 & -7.86$\pm$0.14 \\ 
             & Upper Sco & 1.33$\pm$0.29 & -9.17$\pm$0.21 \\ 
            \hline
            $M_*-M\mathrm{_{disk}}$ & Ophiuchus & 1.07$\pm$0.15 & -2.55$\pm$0.12 \\ 
             & Lupus & 1.48$\pm$0.18 & -2.03$\pm$0.14 \\ 
             & (only members) & 1.38$\pm$0.19 & -2.06$\pm$0.14 \\ 
             & Cha-I & 1.24$\pm$0.17 & -2.46$\pm$0.12 \\ 
             & Upper Sco & 1.22$\pm$0.19 & -3.16$\pm$0.15 \\ 
            \hline
            $M_*-M\mathrm{_{disk}}$ & Ophiuchus & 1.27$\pm$0.28 & -2.48$\pm$0.15 \\ 
            ($>$0.1$M_\odot$) & Lupus & 1.39$\pm$0.27 & -2.06$\pm$0.16 \\ 
             & (only members) & 1.39$\pm$0.27 & -2.06$\pm$0.16 \\ 
             & Cha-I & 1.13$\pm$0.27 & -2.48$\pm$0.13 \\ 
             & Upper Sco & 1.51$\pm$0.23 & -3.03$\pm$0.16 \\ 
            \hline
            $M\mathrm{_{disk}}-\dot{M}\mathrm{_{acc}}$ & Ophiuchus & 0.75$\pm$0.13 & -6.33$\pm$0.41 \\ 
             & Lupus & 0.81$\pm$0.1 & -6.86$\pm$0.31 \\ 
             & (only members) & 0.73$\pm$0.11 & -7.03$\pm$0.32 \\ 
             & Cha-I & 0.95$\pm$0.12 & -6.0$\pm$0.39 \\ 
             & Upper Sco & 0.9$\pm$0.14 & -6.47$\pm$0.58 \\ 
            \hline
            $M\mathrm{_{disk}}-\dot{M}\mathrm{_{acc}}$ & Ophiuchus & 0.5$\pm$0.11 & -6.82$\pm$0.34 \\ 
            ($>$0.1$M_\odot$) & Lupus & 0.57$\pm$0.12 & -7.39$\pm$0.34 \\ 
             & (only members) & 0.57$\pm$0.12 & -7.39$\pm$0.34 \\ 
             & Cha-I & 0.71$\pm$0.14 & -6.45$\pm$0.41 \\ 
             & Upper Sco & 0.92$\pm$0.16 & -6.37$\pm$0.63 \\ 
            \hline 
        \end{tabular}
    \end{center}
    \label{tab:tab_slopes}
\end{table}

\end{appendix}

\end{document}